\documentclass[12pt,a4paper]{article}
\usepackage[truedimen,margin=30mm]{geometry} 

\usepackage{mathrsfs}
\usepackage{amssymb}
\usepackage{amsmath}
\usepackage{ascmac}
\usepackage[dvips]{graphicx}
\usepackage{natbib}
\usepackage{setspace}
\usepackage{times}
\usepackage{bm}  

\usepackage{color}

\usepackage{algorithmic}
\usepackage{algorithm}
\usepackage{siunitx}

\usepackage{titlesec}
\titleformat*{\section}{\large\bfseries}
\titleformat*{\subsection}{\it}

\newtheorem{prp}{Proposition}

\title{{\bf Propensity Patchwork Kriging for Scalable Inference on Heterogeneous Treatment Effects}\footnote{\today}}

\date{}

\begin{document}

\maketitle
\doublespacing

\vspace{-1.5cm}
\begin{center}
{\large Hajime Ogawa$^1$ and Shonosuke Sugasawa$^2$}

\medskip

\medskip
\noindent
$^1$Graduate School of Economics, Keio University\\
$^2$Faculty of Economics, Keio University\\
\end{center}

\vspace{0.5cm}
\begin{center}
{\bf \large Abstract}
\end{center}

Gaussian process-based models are attractive for estimating heterogeneous treatment effects (HTE), but their computational cost limits scalability in causal inference settings. In this work, we address this challenge by extending Patchwork Kriging into the causal inference framework. Our proposed method partitions the data according to the estimated propensity score and applies Patchwork Kriging to enforce continuity of HTE estimates across adjacent regions. By imposing continuity constraints only along the propensity score dimension, rather than the full covariate space, the proposed approach substantially reduces computational cost while avoiding discontinuities inherent in simple local approximations. The resulting method can be interpreted as a smoothing extension of stratification and provides an efficient approach to HTE estimation. The proposed method is demonstrated through simulation studies and a real data application.

\vspace{-0cm}


\newpage

\section{Introduction}
The heterogeneous treatment effect (HTE) is an important concept in causal inference. In many situations such as medicine, marketing and public policy, analysts are interested in knowing the treatment effect at the individual level rather than the entire population or a specific group. 
Estimating HTE is a challenging problem and several studies have approached this problem using machine learning methods. For example,
\cite{louizos2017causaleffectinferencedeep} proposed a neural network-based method to adjust for unobserved confounders. In the Bayesian framework, \citet{hahn2019bayesianregressiontreemodels} proposed Bayesian Causal Forest (BCF), a tree-based model for estimating treatment effects at the individual level. Bayesian methods are advantageous as they allow for the evaluation of uncertainty in HTE estimates. When considering the use of causal inference for individual-level decision-making, the ability to evaluate the uncertainty in HTE estimates is crucial.

In this paper, we focus on methods based on the Gaussian process (GP). GP models are a representative choice for HTE estimation in Bayesian machine learning methods. \cite{NIPS2017_6a508a60} proposed the \textit{Multi-task Gaussian Process} for factual and counterfactual responses using GP models with a kernel that allows for mutual information sharing. \cite{horii2023uncertaintyquantificationheterogeneoustreatment} proposed using GP priors in a partially linear model and proved that the posterior of the HTE concentrates around the true one, as the sample size approaches infinity. GP models are highly flexible methods, allowing them to capture nonlinear and complex treatment effects. Furthermore, their smoothness enables the estimation of treatment effects even in situations where the overlap between treatment and control groups is weak \citep{zhu2022addressingpositivityviolationscausal}.
A major drawback of Gaussian process (GP) models is their high computational cost. For a sample size $n$, the computational complexity is generally on the order of $O(n^3)$. Significant research has been dedicated to addressing this issue. One representative approach is the \textit{Sparse Gaussian Process} \citep{NIPS2005_4491777b}, which approximates the covariance matrix using a set of inducing points (pseudo-inputs). \cite{pmlr-v5-titsias09a} further advanced this by proposing a variational inference framework for optimal selection of these inducing points. Another popular category is low-rank approximation. This method approximates the covariance matrix $K$ using a low-rank matrix product $QQ^T$ \citep{Rasmussen2006Gaussian}. Within this approach, the Nyström method \citep{NIPS2000_19de10ad}, which uses a subset of data to compute the top eigenvalues for the approximation, is a standard technique. More recently, \cite{Datta02042016} proposed the \textit{Nearest Neighbor Gaussian Processes}, which has been extended to various other settings \citep{finley2018efficientalgorithmsbayesiannearest, DBLP:journals/corr/abs-2202-01694}.

In this paper, we improve the computational efficiency of GP models for the HTE estimation by introducing \textit{Patchwork Kriging}. The method is based on local approximation, reducing the computational cost by dividing the data into multiple regions. This approach not only reduces computational costs but also allows for region-specific hyperparameter tuning. The main idea of Patchwork Kriging is to obtain a continuous prediction across the boundaries by introducing a latent variable considering the difference between two local models. We propose estimating the propensity score $e(\bm{x})=\mathbb{P}(T=1|\bm{x})$ in advance and using it as the sole partitioning variable in this phase. Our approach extends Patchwork Kriging in that it partitions the data using a one-dimensional variable derived from the features. This idea makes the boundaries between subregions more explicit and helps to address potential imbalances in sample sizes across these regions. Furthermore, our model can be interpreted as a natural extension of stratification in causal inference. In the causal inference framework, stratification methods—which divide the entire dataset into subgroups based on propensity scores and analyze effects within each group—are widely used. This approach has also been extended to the Bayesian framework \citep{https://doi.org/10.1002/sim.3460, orihara2024bayesianbasedpropensityscoresubclassification}. 
However, from the perspective of HTE estimation, the results of these methods are discontinuous at the boundary points. Our proposed method mitigates these unnatural discontinuities inherent in conventional methods. 
A key practical motivation underlying our approach is that, for scalable computation, continuity constraints can be imposed along an interpretable one-dimensional balancing score rather than directly over the full covariate space. 
This construction does not imply that smoothness along the propensity score is a substitute for smoothness in the original covariate space. 
Rather, it provides a computationally tractable stitching structure for local GP approximations, avoiding the geometric and computational challenges of constructing pseudo-observations in high-dimensional input domains.
We experimentally demonstrate the efficacy of our proposed method. Simulation studies show that our method is computationally more efficient than standard GP-based baselines and achieves higher accuracy than simple local approximations. Furthermore, in a large-scale real-world application ($N=7752$), the proposed method substantially reduced the computation time compared with the baseline global GP.

The remainder of this paper is organized as follows. 
In Section 2, we introduce the notation and framework for HTE estimation used in this paper and review the methodology of Patchwork Kriging. 
Section 3 describes the proposed method, including the full algorithm and the derivation of the target posterior distribution. 
Section 4 reports simulation studies comparing the proposed method with existing methods. 
Section 5 presents an empirical application using medical expenditure data. 
Finally, Section 6 concludes the paper with summary comments.

\section{Preliminary}
\subsection{Notation and framework}
This section defines the notation and describes the framework of our model. We consider binary assignment $T\in \{0,1\}$. The dataset consists of $N$ \textit{i.i.d.} samples $\{\bm{x}_n, Y_n, T_n \}_{n=1,\dots, N}$. Let $\bm{X}$ be a vector of covariates, and $\bm{x}$ be a realization of $\bm{X}$. Our method is based on the \textit{potential outcome framework} \citep{Rubin:1974}. To estimate HTE, we make the following standard assumptions. The first is the \textit{strong ignorability condition}; $\{Y^{0},Y^{1}\} \mathop{\perp\!\!\!\!\perp} T|\bm{X}$. 
This assumption guarantees independence between potential outcomes and treatment assignment given pre-treatment covariates $\bm{X}$. The second is the \textit{positivity condition}; $0 < \mathbb{P}(T=1|\bm{x}) < 1$. Here, our main interest is estimating HTE on test data.

Our proposed model is a type of \textit{partially linear model} \citep{34695649-b87d-3b87-a786-c94fff3cdc21}. The partially linear model is 
\begin{align*}
    Y = \theta T + f(\bm{X}) + \epsilon, \quad \epsilon \sim \mathcal{N}(0,s_{\epsilon}^{-1})
\end{align*}
In this model, the baseline of the outcome $Y^{0}$ is modeled by a non-linear function $f(\cdot)$ and HTE is represented by a constant parameter $\theta$. In this modeling framework, any suitable type of model can be adapted to $f(\cdot)$. 
Building on this, \cite{horii2023uncertaintyquantificationheterogeneoustreatment} proposed a variant type of the partially linear model
\begin{align*}
    Y = \theta(\bm{X}) T + f(\bm{X}) + \epsilon, \quad 
    \epsilon \sim \mathcal{N}(0,s_{\epsilon}^{-1})
\end{align*}
In this model, a non-linear function is also adapted to $\theta(\cdot)$, allowing the model to capture complex, individual-level treatment effects. Furthermore, they proposed using zero-mean GP prior to both $\theta(\cdot)$ and $f(\cdot)$.
\begin{align*}
\theta(\cdot) \sim \mathcal{GP}(0, C(.,.;\bm{\omega}_{\theta}), \quad 
f(\cdot) \sim \mathcal{GP}(0, C(.,.;\bm{\omega}_{f}))
\end{align*}

Our proposed model builds upon this formulation. 
It is worth noting that, in the binary treatment setting, the partially linear representation with a covariate-dependent treatment effect is not a restrictive structural assumption on the conditional mean function. 
Indeed, for any pair of conditional mean functions $\mu_0(x)=E(Y\mid X=x,T=0)$ and $\mu_1(x)=E(Y\mid X=x,T=1)$, we can write
$E(Y\mid X=x,T)=f(x)+T\theta(x)$
by setting $f(x)=\mu_0(x)$ and $\theta(x)=\mu_1(x)-\mu_0(x)$. 
Thus, $\theta(x)$ directly corresponds to the conditional treatment effect under the standard causal assumptions. 
This decomposition is also useful for modeling and computation, because it allows us to assign separate GP priors and kernel hyperparameters to the baseline response function $f(\cdot)$ and the HTE function $\theta(\cdot)$.

\subsection{Review of Patchwork Kriging}
This section reviews \textit{Patchwork Kriging}. First, we consider partitioning the input space of $X$ into $K$ local regions, denoted as $\{\Omega_1, \dots, \Omega_K\}$, with corresponding boundaries $\{\Gamma_{1,2}, \dots, \Gamma_{K-1,K}\}$. 
Let $f_k$ be the $k$-th local model which takes an input $x_k \in \Omega_k$. 
The local models $f_k$ are assumed to be mutually independent Gaussian processes with zero means and covariance functions $c_k$.

The objective of this method is to obtain a continuous prediction across the boundaries. Patchwork Kriging considers the difference between two local models on a shared boundary. Define the latent variable $\bm{\delta}$ as follows;
\begin{align*}
\bm{\delta}_{k,l}(x) = f_k(x) - f_l(x) \ \ {\rm for} \ \ x\in\Gamma_{k,l}.
\end{align*}
To avoid duplication, we assume $k < l$.
Note that $\bm{\delta}_{k,l}$ is the difference between two local models of neighboring regions $\Omega_k$ and $\Omega_l$ where these areas are in contact $\Gamma_{k,l}$. The key idea of Patchwork Kriging is to derive the posterior predictive distribution by constraining these differences to be zero (i.e., $\bm{\delta}=0$). 
After imposing this constraint, initially independent models will be dependent. In practice, since there are infinitely many points on each boundary $\Gamma_{k,l}$, the constraint is enforced on $B$ points randomly sampled from the boundary.

Since our local models are GPs, their difference $\bm{\delta}$ is also a GP. It has a zero-mean and a covariance function of $c_k(\cdot,\cdot)+c_l(\cdot,\cdot)$. 
From the definition of $\bm{\delta}_{k,l}$, the covariance between $\bm{\delta}_{k,l}(x_1)$ and $f_j(x_2)$ is given by
\begin{equation*}
 \begin{split}
\mathbb{C}{\rm ov}(\bm{\delta}_{k,l}(x_1), f_j(x_2)) &= \mathbb{C}{\rm ov}(f_k(x_1) - f_l(x_1), f_j(x_2)) \\
    &= \mathbb{C}{\rm ov}(f_k(x_1), f_j(x_2)) - \mathbb{C}{\rm ov}(f_l(x_1), f_j(x_2))
 \end{split}
\end{equation*}
If $k=j$, $\mathbb{C}{\rm ov}(f_k(x_1), f_j(x_2))=c_k(x_1,x_2)$, otherwise take $0$. So,
\begin{align}
\label{delta_f_cov_rule}
    \mathbb{C}{\rm ov}(\bm{\delta}_{k,l}(x_1), f_j(x_2)) = \left\{
\begin{array}{ll}
c_k(x_1,x_2) & k=j \\
-c_l(x_1,x_2) & l=j \\
0 & {\rm otherwise}.
\end{array}
\right.
\end{align}
If the local model $f_j$ does not share the boundary $\Gamma_{k,l}$, the covariance takes $0$. 
Similarly, the covariance between $\bm{\delta}_{k,l}$ and $\bm{\delta}_{u, v}$ is 
\begin{equation}
\label{delta_delta_cov_rule}
    \mathbb{C}{\rm ov}(\bm{\delta}_{k, l}(x_1), \bm{\delta}_{u, v}(x_2)) =
    \left\{
    \begin{array}{cccccc}
         c_k(x_1, x_2)&  k=u, l\neq v\\
         c_l(x_1, x_2)&  l=v, k\neq u\\
         -c_k(x_1, x_2)&  k=v, l\neq u\\
         -c_l(x_1, x_2)&  l=u, k\neq v\\
         c_k(x_1, x_2) + c_l(x_1, x_2) &  k=v, l=u\\
         0 & otherwise.
    \end{array}
    \right.
\end{equation}
Let $x^*$  be a new input, and we assume $x^*\in\Omega_k$. Additionally, we define $\bm{\delta}$ as the collection of $\bm{\delta}_{k,l}$.
The joint distribution of $y$, $\bm{\delta}$ and $f_k(x^*)$ is 
\begin{align*}
    \begin{bmatrix}
    f_k^*\\
    y\\
    \bm{\delta}
    \end{bmatrix} \sim \mathcal{N} \left(
    \begin{bmatrix}
        0\\0\\0
    \end{bmatrix},
    \begin{bmatrix}
        C_{**}& C_{*D} & C_{*\bm{\delta}}\\
        C_{D*}& C_{DD} & C_{D\bm{\delta}}\\
        C_{\bm{\delta}*}& C_{\bm{\delta} D} & C_{\bm{\delta}\bm{\delta}}
    \end{bmatrix}
    \right)
\end{align*}
Here, we use a local approximation, so $C_{DD}$ is a block diagonal matrix. 
Also, from \eqref{delta_delta_cov_rule}, $C_{\bm{\delta},*}$ is a sparse matrix. For similar reasons, other covariance matrices are sparse matrices too. 

Following the standard equations for GP regression, the posterior predictive distribution is Gaussian with the following mean and covariance: 
\begin{align}
\label{PK_pred_mean}
    \mathbb{E}[f_k^*|y,\bm{\delta}] = (C_{*D}-C_{*\bm{\delta}}C_{\bm{\delta}\bm{\delta}}^{-1}C_{D\bm{\delta}}^T)(C_{DD}-C_{D\bm{\delta}}C_{\bm{\delta}\bm{\delta}}^{-1}C_{D\bm{\delta}}^T)^{-1}y,
\end{align}
and 
\begin{equation}
\label{PK_pred_cov}
 \begin{split}
\mathbb{C}{\rm ov}(f_k^*|y,\bm{\delta}) &= 
C_{**} - C_{*\bm{\delta}}C_{\bm{\delta}\bm{\delta}}^{-1} C_{\bm{\delta}*}^T \\
& - (C_{*\bm{\delta}}C_{\bm{\delta}\bm{\delta}}^{-1}C_{D\bm{\delta}}^T)(C_{DD}-C_{D\bm{\delta}}C_{\bm{\delta}\bm{\delta}}^{-1}C_{D\bm{\delta}}^T)^{-1}(C_{D*}-C_{D\bm{\delta}}C_{\bm{\delta}\bm{\delta}}^{-1}C_{\bm{\delta}*}^T).
 \end{split}
\end{equation}

In Patchwork Kriging, the total computational cost is lower than that of a standard GP. 
When calculating \eqref{PK_pred_mean} or \eqref{PK_pred_cov}, the inversion term $(C_{DD}-C_{D\bm{\delta}}C_{\bm{\delta}\bm{\delta}}^{-1}C_{D\bm{\delta}}^T)^{-1}$ dominates the overall computational cost. The total cost is $O(KM^3 + KBM^2 + d_f^3B^3K)$, where $M$ is the sample size in each region and $d_f$ is the number of neighboring regions. Typically, $M \simeq N/K$, so computational cost decreases as the number of partitions $K$ increases.

\section{Scalable Inference on HTE}

A key contribution of our method is an extension of Patchwork Kriging in which continuity constraints are imposed only along the propensity score dimension, rather than over the full covariate space. 
We refer to this approach as propensity Patchwork Kriging (PPK). 
We then employ PPK within the partially linear model for HTE estimation.
The procedure consists of three main steps. First, we divide the data by propensity score and tune hyperparameters of GP priors in each region. Second, we generate pseudo inputs and constrain their propensity scores to match the boundary values. 
Finally, we derive the posterior predictive distribution by conditioning the boundary pseudo-observations to be equal to zero.

\subsection{Dividing data using propensity score}
To adapt Patchwork Kriging to our model, we first divide the data. 
In causal inference, a natural strategy for dividing data is to use propensity score $e(\bm{x})=\mathbb{P}(T=1|\bm{x})$, motivated by propensity score subclassification for causal effect estimation \citep{rosenbaum1983central}. 
There are two advantages to this method. First, by dividing data using a one-dimensional variable, the boundaries between divided regions become clearer. 
If we divide data by multiple variables, the connections would be more complex and implementation of Patchwork Kriging would be more difficult as it requires a larger number of pseudo-observations under a larger number of boundaries between regions. 
Pseudo-observations act as constraints on the model, and thus reducing their number is preferable.
Also, by dividing with multiple variables, a special method is needed to avoid creating regions with small sample size. \cite{park2018patchworkkriginglargescalegaussian} used a spatial tree to construct regions with approximately equal sample sizes.
However, by dividing with a one-dimensional variable, such special methods are not needed and we can divide data more easily.

Second, by using propensity score to divide data, our method can be interpreted as a natural extension of stratification methods in the causal inference framework.
Propensity score is the most popular \textit{balancing score} that constrains $\bm{X}\mathop{\perp\!\!\!\!\perp} T | e(\bm{X})$, and the average treatment effect can be identified by combining strong ignorability. 
For this reason, there are many methods that stratify the entire dataset into subgroups and analyze in each group. 
In HTE estimation, results of these methods are discontinuous at the boundary points. 
Our proposed method mitigates these artificial discontinuities in conventional stratification-based approaches.

\subsection{Hyperparameter tuning}

In our proposed method, we optimize the hyperparameters of the kernels for the treatment effect function $\theta(\cdot)$ and the baseline function $f(\cdot)$, as well as the noise parameter $s_\epsilon$, independently within each local region $\Omega_k$. 
Specifically, we employ the radial basis function (RBF) kernel, parameterized by $\gamma_\theta$ and $\gamma_f$, defined as $C(\bm{x}, \bm{y}) = \exp(- \gamma \|\bm{x} - \bm{y}\|^2)$.
This local optimization allows the model to adapt to region-specific characteristics, such as varying degrees of smoothness or noise levels.

For each local region $\Omega_k$, we maximize the marginal log-likelihood $L_k(\gamma_\theta, \gamma_f, s_\epsilon)$, defined as
\begin{equation}
\label{log_likelihood}
\begin{split}
\mathcal{L}_k(\gamma_\theta, \gamma_f, s_\epsilon) 
= \log \mathcal{N}(\bm{y}_k | \bm{0}, \bm{V}_k), \quad 
\bm{V}_k = \text{diag}(\bm{t}_k) \bm{C}_{\theta, k} \text{diag}(\bm{t}_k) + \bm{C}_{f, k} + \frac{1}{s_\epsilon} \bm{I}_{k},
\end{split}
\end{equation}
where $\bm{y}_k$ and $\bm{t}_k$ are the vectors of outcomes and treatment assignments for the samples in region $\Omega_k$, respectively. $\bm{C}_{\theta, k}$ and $\bm{C}_{f, k}$ denote the covariance matrices computed using the RBF kernel within the region. 
To maximize \eqref{log_likelihood}, we employ a hybrid approach combining grid search and gradient-based optimization. Because the marginal likelihood surface is often highly non-convex with respect to the smoothness parameters ($\gamma_\theta$ and $\gamma_f$), purely gradient-based optimization is prone to getting trapped in local optima. Therefore, following a similar strategy to \citet{horii2023uncertaintyquantificationheterogeneoustreatment}, we perform a grid search over $\gamma_\theta$ and $\gamma_f$.
For each combination of $\gamma_\theta$ and $\gamma_f$, in the grid, we optimize the noise parameter $s_\epsilon$ using a gradient-based method. While \citet{horii2023uncertaintyquantificationheterogeneoustreatment} uses Newton's method for this optimization, it requires computing the full Hessian matrix, which severely limits scalability for large datasets. To overcome this computational bottleneck and maintain scalability, we instead employ the L-BFGS-B algorithm, a quasi-Newton method that efficiently approximates the Hessian and naturally handles the boundary constraints of the parameters.

Finally, we select the set of hyperparameters $(\gamma_\theta, \gamma_f, s_\epsilon)$ that maximizes the marginal log-likelihood for each subregion. This localized tuning phase is highly computationally efficient because the optimization for each region can be executed completely in parallel. Consequently, the total computational time is significantly lower than that required for a standard global GP model.
In this paper, we set the search grids for both $\gamma_\theta$ and $\gamma_f$ to range from 0.1 to 5.0 with an increment of 0.2.

\subsection{Generation of pseudo-observations}

In this step, we generate pseudo-observations at each boundary to stitch the local GP models. 
Since the data are partitioned into $K$ regions based on the one-dimensional propensity score, there are $K-1$ boundaries. 
We generate $B$ pseudo-observations at each boundary, resulting in a total of $B(K-1)$ additional inputs. 
Crucially, the propensity score of each pseudo-observation must exactly equal the corresponding boundary value under the fitted propensity model.
To ensure that the generated pseudo-observations remain realistic and close to the actual data manifold, we propose a deterministic procedure based on observed cross-boundary nearest-neighbor pairs. 

For clarity, we first describe the procedure when all covariates are continuous. 
Let $c_k$ denote the boundary value of the fitted propensity score between the $k$th and
$(k+1)$th regions, and let $\mathcal{D}_k$ and $\mathcal{D}_{k+1}$ be the sets of observed covariates in the two adjacent regions. We generate $B$ pseudo-observations for this
boundary.
Suppose that the propensity score is estimated by the logistic regression model, $\mathrm{logit}\{\widehat e(x)\}=\widehat\beta_0+x^\top\widehat\beta$ with $\widehat\beta=(\widehat\beta_1,\ldots,\widehat\beta_p)$ and ${\rm logit}(u)=\log(u/(1-u))$. 
Let $\delta>0$ be a threshold for the absolute value of the regression coefficients, and define $\mathcal{P}_{\mathrm{adj}}$ as the ordered set of covariate indices satisfying
$|\widehat\beta_\ell|>\delta$ among $\ell=1,\ldots,p$, sorted in decreasing order of $|\widehat\beta_\ell|$. 
In our implementation, we use $\delta=0.1$ by default.
For the boundary between the $k$th and $(k+1)$th regions, pseudo-observations are
generated as follows.

\begin{enumerate}
\item 
Initialize the number of generated pseudo-observations as $b=0$.

\item 
Select $x_k\in\mathcal{D}_k$ and $x_{k+1}\in\mathcal{D}_{k+1}$ whose fitted propensity
scores are closest to the boundary value $c_k$ from each side, namely, $x_k={\rm argmin}_{x\in \mathcal{D}_k}|\widehat{e}(x)-c_k|$ and $x_{k+1}={\rm argmin}_{x\in \mathcal{D}_{k+1}}|\widehat{e}(x)-c_k|$.

\item 
For each $\ell \in P_{\rm adj}$, taken sequentially in the prescribed order, construct a pseudo-observation $x^\ast$ as follows.
First, initialize the covariates other than the $\ell$th coordinate by the midpoint of the selected cross-boundary pair,
$x^\ast_{-\ell}=(x_{k,-\ell}+x_{k+1,-\ell})/2$. 
This midpoint is used as a local interpolation point between two observed units located near the boundary. 
Then determine the remaining coordinate $x^\ast_\ell$ so that $\widehat e(x^\ast)=c_k$. 
Under the logistic propensity score model, this gives
$x^\ast_\ell=
\{{\rm logit}(c_k)-\widehat\beta_0-x^{\ast\top}_{-\ell}\widehat\beta_{-\ell}\}/\widehat\beta_\ell$.
After generating $x^\ast$, update $b\leftarrow b+1$. 
If $b=B$, stop the procedure and go to the next boundary. 
Otherwise, continue to the next index in $P_{\rm adj}$.

\item 
If $b<B$, update the set of covariates as $\mathcal{D}_k \leftarrow  \mathcal{D}_k\setminus \{x_k\}$ and $\mathcal{D}_{k+1} \leftarrow  \mathcal{D}_{k+1}\setminus \{x_{k+1}\}$, namely, remove the selected pair $(x_k,x_{k+1})$ from $\mathcal{D}_k$ and $\mathcal{D}_{k+1}$, and then return to Step 2; otherwise the algorithm is terminated. 
\end{enumerate}

When discrete or categorical covariates are included, the above procedure can be modified in a simple way.
Such covariates are excluded from $\mathcal{P}_{\mathrm{adj}}$, since their values cannot be continuously adjusted to satisfy the boundary condition $\widehat e(x^*)=c_k$.
When constructing $x^*_{-\ell}$ for an adjustment coordinate $\ell$, continuous covariates are averaged as described above, whereas discrete or categorical covariates are assigned one of the two observed values from the selected cross-boundary pair.
Specifically, we alternately use the value from the $k$th region and that from the $(k+1)$th region across iterations.
This prevents the generated pseudo-observations from taking invalid intermediate values and keeps them within the support of the observed covariates.

%
\subsection{Predictive distribution of HTE}

In this phase, we construct the prior covariance matrix following the rules of \cite{park2018patchworkkriginglargescalegaussian}.
We consider the partially linear model, $ Y = \theta(\bm{X}) T + f(\bm{X}) + \epsilon$. To derive the predictive distribution, we will use the following prior. 
\begin{equation}
\label{prior}
    p(\bm{\Theta}) = p \begin{pmatrix} \bm{\theta}_n\\ \tilde{\bm{\theta}}_m\\\bm{f}_n\\ \bm{\delta} \end{pmatrix} 
    =
    \mathcal{N}\left( \bm{0}, \begin{bmatrix}
        \Sigma_{\bm{\theta}_n, \bm{\theta}_n} & \Sigma_{\bm{\theta}_n, \tilde{\bm{\theta}}_m} & \bm{0} & \Sigma_{\bm{\theta}_n, \bm{\delta}} \\
        \Sigma_{\tilde{\bm{\theta}}_m, \bm{\theta}_n} & \Sigma_{\tilde{\bm{\theta}}_m, \tilde{\bm{\theta}}_m} & \bm{0} & \Sigma_{\tilde{\bm{\theta}}_m, \bm{\delta}} \\
        \bm{0} & \bm{0} & \Sigma_{\bm{f}_n, \bm{f}_n} & \bm{0}\\
        \Sigma_{\bm{\delta}, \bm{\theta}_n} & \Sigma_{\bm{\delta}, \tilde{\bm{\theta}}_m} & \bm{0} & \Sigma_{\bm{\delta}, \bm{\delta}}
    \end{bmatrix}\right).
\end{equation}
$\bm{\theta}_n$ and $\tilde{\bm{\theta}}_m$ are vectors of HTE values evaluated at the training and test input points, respectively. 
$\delta$ is the vector of boundary pseudo-observations defined as
\begin{equation*}
    \bm{\delta} = \begin{pmatrix} 
                   \bm{\delta_{1,2}} \\
                   \bm{\delta_{2,3}} \\\vdots \\ \bm{\delta_{K-1,K}}
             \end{pmatrix}  = 
    \begin{pmatrix}
        \theta^2({\bm{X}^{\rm pseudo}_{1,1,2}}) - \theta^1(\bm{X}^{\rm pseudo}_{1,1,2}) \\ \vdots \\ \theta^K(\bm{X}^{\rm pseudo}_{B,K-1,K}) - \theta^{K-1}(\bm{X}^{\rm pseudo}_{B, K-1,K})
    \end{pmatrix}
\end{equation*}
Here, $\theta_k(\cdot)$ denotes the HTE function in the $k$th region, and $\delta$ is the vector of these boundary differences.
Covariance matrices of HTE and nuisance parameter functions  $\Sigma_{\bm{\theta}_n, \bm{\theta}_n}$,$\Sigma_{\bm{\theta}_n, \tilde{\bm{\theta}}_m}$ and $\Sigma_{\tilde{\bm{\theta}}_m, \tilde{\bm{\theta}}_m}$ are block diagonal matrices. 
For example, if $K=3$, $\Sigma_{\bm{\theta}_n, \bm{\theta}_n}$ is
\begin{equation*}
    \Sigma_{\bm{\theta}_n, \bm{\theta}_n} = 
    \begin{bmatrix}
        \Sigma_{\bm{\theta}^1, \bm{\theta}^1}  & \bm{0} &  \bm{0}\\
        \bm{0} & \Sigma_{\bm{\theta}^2, \bm{\theta}^2}  & \bm{0}\\
        \bm{0}& \bm{0} & \Sigma_{\bm{\theta}^3, \bm{\theta}^3} 
    \end{bmatrix}
\end{equation*}
$\Sigma_{\bm{\theta}^1, \bm{\theta}^1}$ is kernel matrix for the training data in 1st region. In this work, we set the off-diagonal blocks to $\bm{0}$, making the entire matrix sparse. $\Sigma_{\bm{\theta}_n, \tilde{\bm{\theta}}_m}$ and $\Sigma_{\tilde{\bm{\theta}}_m, \tilde{\bm{\theta}}_m}$ can be treated similarly.
Because each boundary pseudo-observation $\delta_{k,k+1}$ involves two adjacent regions, some off-diagonal blocks are nonzero.
For example, if $K=4$, it follows from \eqref{delta_f_cov_rule} that $\Sigma_{\bm{\delta}, \bm{\theta}_n}$ is
\begin{equation*}
    \Sigma_{\bm{\delta}, {\bm\theta}_n} = 
    \begin{bmatrix}
        \Sigma_{\bm{\theta^1}, \bm{\delta_{1,2}}} &  \Sigma_{\bm{\theta^2}, \bm{\delta_{1,2}}} & \bm{0} & \bm{0}\\
        \bm{0} &  \Sigma_{\bm{\theta^2}, \bm{\delta_{2,3}}} &
        \Sigma_{\bm{\theta^3}, \bm{\delta_{2,3}}} & \bm{0} \\
        \bm{0} & 
        \bm{0} & \Sigma_{\bm{\theta^3}, \bm{\delta_{3,4}}}  & \Sigma_{\bm{\theta^4}, \bm{\delta_{3,4}}}  \\ 
    \end{bmatrix}.
\end{equation*}
Similarly, covariance matrix $\Sigma_{\bm{\delta}, \bm{\delta}}$ is
\begin{equation*}
    \Sigma_{\bm{\delta}, \bm{\delta}} = 
    \begin{bmatrix}
        \Sigma_{\bm{\delta_{1,2}}, \bm{\delta_{1,2}}} &  \Sigma_{\bm{\delta_{1,2}}, \bm{\delta_{2,3}}} & \bm{0} \\
        \Sigma_{\bm{\delta_{2,3}}, \bm{\delta_{1,2}}} &  \Sigma_{\bm{\delta_{2,3}}, \bm{\delta_{2,3}}} &
        \Sigma_{\bm{\delta_{2,3}}, \bm{\delta_{3,4}}} \\
        \bm{0} & 
        \Sigma_{\bm{\delta_{3,4}}, \bm{\delta_{2,3}}} &
        \Sigma_{\bm{\delta_{3,4}}, \bm{\delta_{3,4}}}\\ 
    \end{bmatrix}.
\end{equation*}

The covariance rule in \eqref{delta_delta_cov_rule} is somewhat involved.
However, in this work, we divide the data by a one-dimensional variable, the propensity score, so we use only four cases in  \eqref{delta_delta_cov_rule}, 
diagonal part ($k=v, l=u$), off-diagonal block ($l=u, k\neq v$) or ($k=v, l \neq u$) and  $\bm{0}$ part.

Based on the prior construction, we derive the posterior predictive distribution of new input $\tilde{\bm{X}}_m$. Our method of derivation is based on \citet{horii2023uncertaintyquantificationheterogeneoustreatment}. To obtain an analytic formulation of the HTE distribution, we follow three steps. 
First, we derive joint distribution of training outcomes $\bm{y}_n$ and parameters $\bm{\Theta} = (\bm{\theta}_n, \tilde{\bm{\theta}}_m, \bm{f}_n, \bm{\delta})$. Second, we compute conditional distribution of $\bm{\Theta}$ given $\bm{y}_n$ and $\bm{\delta}$. Finally, we derive the marginal distribution of $\tilde{\bm{\theta}}_m$.

The joint distribution of $\bm{y}_n$ and $\bm{\Theta}$ can be decomposed into the following formulation.
\begin{equation}
\label{joint-distribution}
    p(\bm{y}_n, \bm{\Theta} | \bm{x}_n, \tilde{\bm{X}}_m) = p(\bm{y}_n | \bm{\theta}_n, \bm{f}_n) p(\bm{\Theta}|\bm{x}_n, \tilde{\bm{X}}_m)
\end{equation}
We have obtained the prior distribution $p(\bm{\Theta}|\bm{x}_n, \tilde{\bm{X}}_m)$ in equation \eqref{prior}. The likelihood $p(\bm{y}_n | \bm{\theta}_n, \bm{f}_n)$ is Gaussian distribution with the mean of $T_n \theta(\bm{x}_n) + f(\bm{x}_n)$ and  the precision of $s_{\epsilon}$. We assume that hyperparameters of Gaussian process and precision parameter $s_{\epsilon}$ have already been learned.
To derive the joint distribution, we define $\Sigma_{\bm{\Theta}, \bm{\Theta}}$ and $\Delta_{\bm{\Theta}, \bm{\Theta}}$ such that $p(\bm{\Theta} | \bm{x}_n, \tilde{\bm{X}}_m) = \mathcal{N}(\bm{0}, \Sigma_{\bm{\Theta}, \bm{\Theta}}), \Delta_{\bm{\Theta}, \bm{\Theta}} =\Sigma_{\bm{\Theta}, \bm{\Theta}}^{-1}$.
Note that $\Sigma_{\bm{\Theta}, \bm{\Theta}}$ is the covariance matrix defined in \eqref{prior} and $\Delta_{\bm{\Theta}, \bm{\Theta}}$ is the precision matrix of $\Sigma_{\bm{\Theta}, \bm{\Theta}}$. 
We can compute $\Delta_{\bm{\Theta}, \bm{\Theta}}$ by using the Schur's complement formula as follows: 
\begin{equation}
\label{prior-precision}
     \Delta_{\bm{\Theta}, \bm{\Theta}}
    =
      \begin{bmatrix}
        \Delta_{\bm{\theta}_n, \bm{\theta}_n} & \Delta_{\bm{\theta}_n, \tilde{\bm{\theta}}_m} & \bm{0} & \Delta_{\bm{\theta}_n, \bm{\delta}} \\
        \Delta_{\tilde{\bm{\theta}}_m, \bm{\theta}_n} & \Delta_{\tilde{\bm{\theta}}_m, \tilde{\bm{\theta}}_m} & \bm{0} & \Delta_{\tilde{\bm{\theta}}_m, \bm{\delta}} \\
        \bm{0} & \bm{0} & \Delta_{\bm{f}_n, \bm{f}_n} & \bm{0}\\
        \Delta_{\bm{\delta}, \bm{\theta}_n} & \Delta_{\bm{\delta}, \tilde{\bm{\theta}}_m} & \bm{0} & \Delta_{\bm{\delta}, \bm{\delta}}
    \end{bmatrix}
\end{equation}
By completing the square, we can derive the precision matrix $\hat{\Delta}$ of joint distribution $p(\bm{y}_n, \bm{\Theta} | \bm{x}_n, \tilde{\bm{X}}_m)$, where the details are provided in the Supplementary Material (Section S1.1). The derivation results are as follows.
\begin{equation}
\label{delta-hat}
\begin{split}
    \hat{\Delta } = 
     & \begin{bmatrix}
        \Delta_{\bm{\theta}_n, \bm{\theta}_n} &  \bm{0} &  \bm{0} &  \bm{0} &  \bm{0}\\
         \bm{0}& \Delta_{\tilde{\bm{\theta}}_m, \tilde{\bm{\theta}}_m} &  \bm{0} &  \bm{0} &  \bm{0}\\
         \bm{0}& \bm{0} & \Delta_{\bm{f}_n, \bm{f}_n} & \bm{0} &  \bm{0}\\
        \bm{0} & \bm{0} &  \bm{0}& \Delta_{\bm{\delta}, \bm{\delta}} &  \bm{0}\\
         \bm{0}& \bm{0} & \bm{0} & \bm{0} &  s_{\epsilon} \bm{I}_n
    \end{bmatrix}  \\
    & \qquad  \qquad  \qquad  \qquad + 
    \begin{bmatrix}
        s_{\epsilon} \bm{T}_n^2 & \Delta_{\bm{\theta}_n, \tilde{\bm{\theta}}_m} & 
        \bm{T}_n^T s_{\epsilon} & \Delta_{\bm{\theta}_n, \bm{\delta}} & -s_{\epsilon} \bm{T}_n\\
        \Delta_{\tilde{\bm{\theta}}_m, \bm{\theta}_n} & \bm{0} & \bm{0} &
        \Delta_{\tilde{\bm{\theta}}_m, \bm{\delta}} & \bm{0}\\
         \bm{T}_n^T s_{\epsilon} & \bm{0} & s_{\epsilon}\bm{I}_n  &
         \bm{0} & -s_{\epsilon}\bm{I}_n \\
         \Delta_{\bm{\bm{\delta}, \theta}_n} & \Delta_{\bm{\delta}, \tilde{\bm{\theta}}_m} &\bm{0} & \bm{0} & \bm{0} \\
         -s_{\epsilon}\bm{I}_n \bm{T}_n &  \bm{0} & -s_{\epsilon}\bm{I}_n & \bm{0} & \bm{0}
    \end{bmatrix}
\end{split}
\end{equation}
We define the inverse matrix of this distribution as $\hat{\Sigma} = \hat{\Delta}^{-1}$. $\hat{\Sigma}$ is covariance matrix of \eqref{joint-distribution}. This procedure is computationally expensive, so we compute the inverse matrix by block decomposition method, where the details are provided in the Supplementary Material (Section S1.2).

Next, we derive the conditional distribution $p(\hat{\bm{\Theta}}| \bm{y}_n, \bm{\delta}, \bm{X}_n, \tilde{\bm{X}}_m)$.
To do so, let  $\hat{\Sigma}$ be a covariance matrix of $p(\bm{\Theta}, \bm{y}_n, \bm{\delta} | \bm{X}_n, \tilde{\bm{X}}_m)$.
$$
\hat{\Sigma} = 
\begin{bmatrix}
    \hat{\Sigma}_{\bm{\Theta}, \bm{\Theta}} & \hat{\Sigma}_{\bm{\Theta}, D} \\
    \hat{\Sigma}_{D, \bm{\Theta}} & \hat{\Sigma}_{D, D}
\end{bmatrix}
$$
where $\hat{\Sigma}_{\bm{\Theta}, \bm{\Theta}}$, $\hat{\Sigma}_{D, D}$ and $\hat{\Sigma}_{\bm{\Theta}, D}$ are the covariance matrices corresponding to the variance of $\hat{\bm{\Theta}} = (\bm{\theta}_n, \tilde{\bm{\theta}}_m, \bm{f}_n)^T$ , the variance of $D = (\bm{\delta},\bm{y}_n)^T$ and 
covariance between $\hat{\bm{\Theta}}$ and $D$ respectively.
Using the standard formulas for conditional Gaussian distributions, the distribution of $\hat{\bm{\Theta}}$ conditioned on $D$ is derived as follows.
\begin{equation}
\label{joined posterior predictive distribution}
p(\hat{\bm{\Theta}}| \bm{y}_n, \bm{\delta}=0, \bm{x}_n, \tilde{\bm{X}}_m) = 
\mathcal{N}(\bm{\mu}_{\bm{\Theta}|D}, \Sigma_{\bm{\Theta}|D}) 
\end{equation}
with $\mu_{\bm{\Theta}|D} =  M D$ and $\Sigma_{\bm{\Theta}|D} =  \hat{\Sigma}_{\bm{\Theta}, \bm{\Theta}} - M \hat{\Sigma}_{D, \bm{\Theta}}$, where 
\begin{align*}
M &= \hat{\Sigma}_{\bm{\Theta}, D} \hat{\Sigma}_{D,D}^{-1} = 
\begin{pmatrix}
        M_{\bm{\theta}_n} \\ M_{\tilde{\bm{\theta}}_m} \\ M_{\bm{f}_n}
\end{pmatrix}, \quad 
D = 
    \begin{pmatrix}
        \bm{\delta} \\ \bm{y}_n
    \end{pmatrix} =
    \begin{pmatrix}
        \bm{0} \\ \bm{y}_n
    \end{pmatrix}
\end{align*}
This distribution is joint posterior given data $\bm{y}_n$ and constraint $\bm{\delta}=0$.  
Note that the posterior mean $\bm{\mu}_{\bm{\Theta}|D}$ is a vector partitioned into three blocks, and the posterior covariance $\Sigma_{\bm{\Theta}|D}$ is $3 \times 3$ block matrix.

Finally, our main target predictive distribution of HTE on the test data $\tilde{\bm{\theta}}_m$ is obtained by marginalizing the joint posterior distribution \eqref{joined posterior predictive distribution} except for $\tilde{\bm{\theta}}_m$. The resulting distribution is
\begin{equation*}
    p(\tilde{\bm{\theta}}_m) = \mathcal{N}(M_{\tilde{\bm{\theta}}_m}D, \ \Sigma_
    {\tilde{\bm{\theta}}_m,\tilde{\bm{\theta}}_m|D})
\end{equation*}
where $\Sigma_{\tilde{\bm{\theta}}_m,\tilde{\bm{\theta}}_m|D}$ is $(2,2)$-block of the $\Sigma_{\bm{\Theta}|D}$ at \eqref{joined posterior predictive distribution}. If we are concerned with the distribution of $\tilde{\bm{\theta}}_m$ only, we calculate only the blocks that are needed in $M$, $\bm{\mu}_{\bm{\Theta}|D}$, and $\Sigma_{\bm{\Theta}|D}$.

\subsection{Selection of $K$ and $B$}
\begin{figure}[H]
\centering
\includegraphics[width=\linewidth]{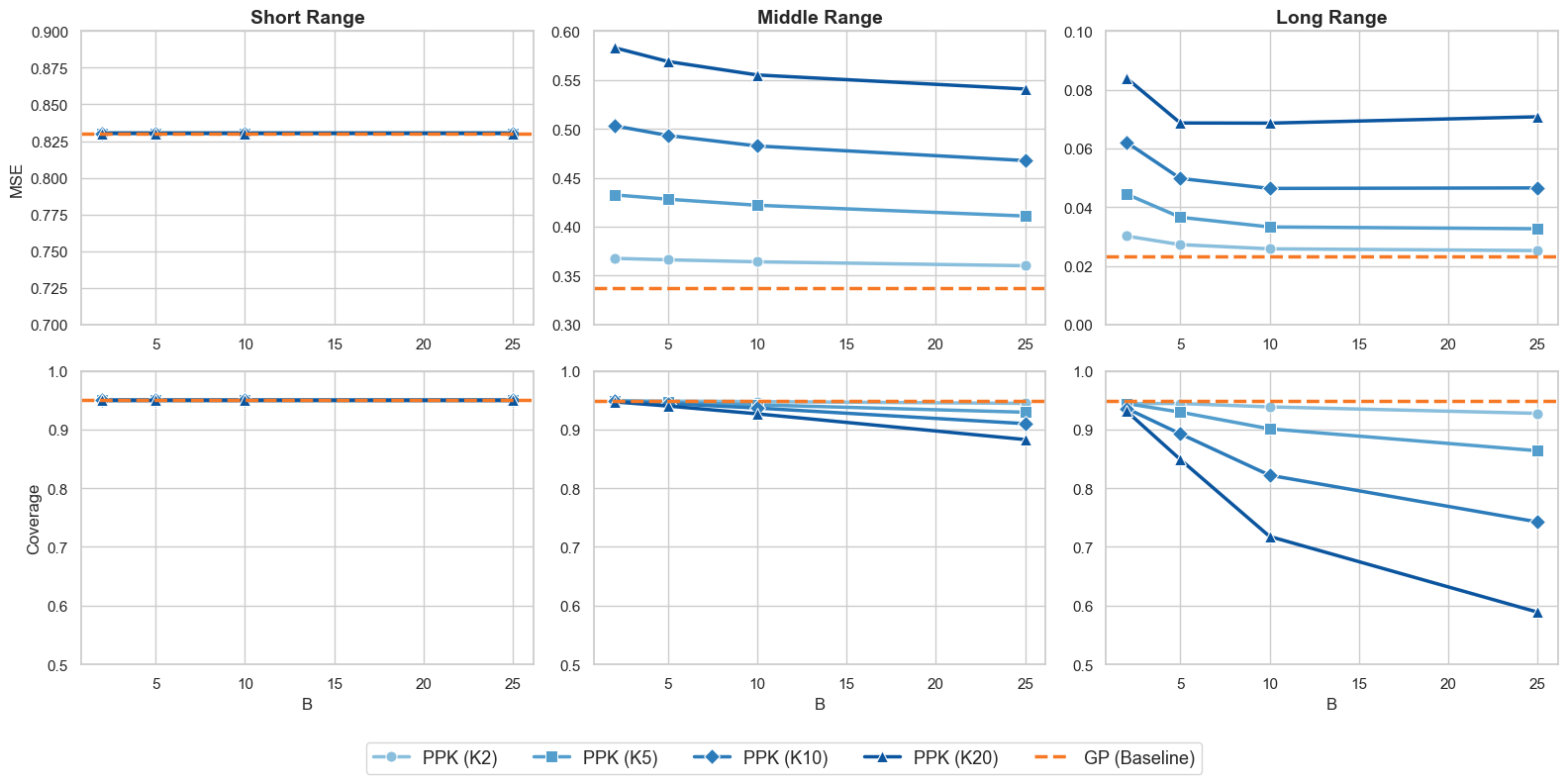}
\caption{Mean squared errors (MSE; top row) and empirical coverage probabilities of 95\% credible intervals (bottom row) for PPK under different numbers of partitions $K$ and boundary pseudo-observations $B$. 
The three columns correspond to short-, middle-, and long-range GP settings. 
The dashed horizontal line indicates the baseline global GP.}
\label{K_B_choise}
\end{figure}

In this section, we discuss the selection of the number of partitions $K$ and the number of pseudo-inputs $B$ on each boundary, and their impact on the performance of PPK.
To this end, we conduct a simple numerical illustration. 
Let $x \in \mathbb{R}^5$ be a confounder vector generated from a multivariate normal distribution, and let $Z \in \{0,1\}$ be a treatment indicator generated from a Bernoulli distribution $\text{Ber}(p=0.5)$. The outcome variable $Y$ is generated by $Y = f(x) + Z\tau(x) + \epsilon$. In this setting, we use GPs with an RBF kernel, where the length-scale is $l \in \{0.1, 1, 5\}$ for $f(\cdot)$ and $\tau(\cdot)$, and the noise term $\epsilon$ is drawn from $\mathcal{N}(0, 1)$. 
Here, we set the sample size to $n=2000$ and assume the true hyperparameters are known. 
The true HTE function $\tau(\cdot)$ has a constant scale across the entire region. Thus, this setting is inherently more advantageous for the baseline global GP model than for PPK.
We compare the standard GP model (with known hyperparameters) and the proposed PPK with $K \in \{2, 5, 10, 20\}$ and $B \in \{2, 5, 10, 25\}$. 
The results of mean squared errors (MSE) and empirical coverage probability of $95\%$ credible intervals are shown in Figure \ref{K_B_choise}.

Figure \ref{K_B_choise} mainly demonstrates the following two points: First, increasing the number of partitions $K$ accelerates the computation speed but degrades the estimation accuracy. The magnitude of this negative impact depends on the GP hyperparameters; in the short-range scenario (e.g., $l=0.1$), the adverse effect on estimation accuracy is minimal.
Second, increasing the number of pseudo-observations $B$ on the boundaries improves the estimation accuracy but negatively affects the empirical coverage. 
This is because conditioning on artificial boundary constraints can make the local models overly confident, leading to an underestimation of the credible intervals. 
This accuracy-coverage trade-off is particularly pronounced in the long-range scenario ($l=5$). Furthermore, the improvement in accuracy from increasing $B$ diminishes beyond a certain point. In the long-range scenario, the improvement appears to plateau at $B=10$. This highlights the inherent limitation of our proposed method, which stitches together local models only at their boundaries.
Based on these results, to achieve sufficient estimation accuracy while maintaining reasonable uncertainty quantification even in long-range cases, it is important to maintain an adequate sample size within each region (e.g., $N_k=N/K\ge 100$) and keep $B$ moderate relative to the sample size in each region (e.g., $N_k/B\ge 10$). 
Therefore, posterior intervals from PPK should be interpreted as approximate uncertainty summaries under the partition-and-stitch approximation, rather than as exact posterior credible intervals from the corresponding full GP model.

\subsection{Posterior consistency of PPK}
\label{sec:consistency}

We provide a theoretical justification for the asymptotic behavior of PPK under fixed $K$.
Let $\tau_0(x)$ denote the true heterogeneous treatment effect.
Suppose that the covariate space is divided into a fixed number of regions $\Omega_1,\ldots,\Omega_K$ according to the propensity score, where $K$ does not depend on $N$.
Let $N_k$ be the number of observations in $\Omega_k$ and assume that $N_k\to\infty$ as $N\to\infty$.
Let $\Pi_{{\rm PPK},N}(\cdot|\boldsymbol{y}_n,\boldsymbol{\delta}=\mathbf{0})$ denote the posterior distribution of $\tau$ obtained by PPK, where $\delta=0$ represents the patchwork boundary constraints.
Let $P_X$ be the marginal distribution of $X$ under the true data-generating process, and define $\|g\|_{L_2(P_X)}=\{\int g(x)^2dP_X(x)\}^{1/2}$.
Let $P_0$ denote the true probability law of the observed data.
The following proposition shows that PPK is posterior consistent when the number of boundary pseudo-observations is asymptotically negligible relative to the sample size, where the technical details are provided in Supplementary Material.

\begin{prp}
\label{prp:ppk_consistency}
Assume the regularity conditions given in the Supplementary Material. 
Suppose that both the number of partitions $K$ and the number of pseudo-observations per boundary $B$ are fixed as $N\to\infty$. 
Then, for every $\epsilon>0$, it holds that
\[
\Pi_{\mathrm{PPK},N}\left\{\tau:
\|\tau-\tau_0\|_{L^2(P_X)}>\epsilon
\mid \boldsymbol y_N,\boldsymbol\delta=\boldsymbol 0
\right\}
\to 0
\]
in $P_0$-probability as $N\to\infty$.
\end{prp}

The fixed-$K$ and fixed-$B$ regime reflects the standard use of PPK in which a finite number of boundary constraints is introduced to stitch local GP models, while the amount of observed data increases. 
Under this regime, the patchwork conditioning term is finite-dimensional and has only a bounded log-scale contribution. 
Therefore, it does not change the large-sample posterior target of the local GP partially linear models. 
The boundary constraints should thus be viewed as a finite-sample smoothing device that reduces artificial discontinuities across adjacent local GP models.

\section{Experiments}

In this section, we conduct simulation studies to evaluate the performance of our proposed model. The objective of the experiments is to estimate the HTE on synthetic data. 
The data generating processes are adapted from \citet{nie2020quasioracleestimationheterogeneoustreatment}. 
We conducted simulation studies on four scenarios. 
In all settings, the covariates are set as $\bm{x}\in \mathcal{R}^6$  and data-generating processes are complicated. 
We set the training sample size to $N\in\{200,500,1000\}$ and the test sample size to $M=500$ in all cases.
The entire process, from data generation to inference, was repeated 100 times in each scenario. Further details are provided in the Supplementary Material (Section S2).
Here, we report the results for two representative scenarios, Setups A and C.
In Setup A, the true HTE function and the true propensity score both depend on $X_1$ and $X_2$, whereby data dividing using estimated propensity score is expected to be effective.
In Setup C, the true HTE function is constant at $1$, so the dividing method itself is not considered to be effective.

In this study, we use the partially linear model with GP as the baseline method. 
Furthermore, we compare the simple local approximation of the baseline model.
Here, the data-partitioning method and hyperparameter tuning procedure for the local approximation are identical to those used in our proposed method.
When the data are divided to $K$ regions, we use common estimated propensity scores in the local approximation. We set the number of regions to $K \in \{2, 5, 10\}$. Based on the findings in Section 3.5, we carefully control the number of pseudo-observations $B$ generated at each boundary to ensure that the ratio of the regional sample size $N_k$ to $B$ is at least 10. 
Specifically, depending on the total sample size $N$, we configure the combinations of $(K, B)$ as follows: for $N=200$, $(2, 5)$, $(5, 2)$, and $(10, 1)$; for $N=500$, $(2, 10)$, $(5, 5)$, and $(10, 2)$; and for $N=1000$, $(2, 10)$, $(5, 10)$, and $(10, 5)$. As demonstrated in Section 3.5, increasing $B$ beyond 10 yields negligible improvements in estimation accuracy; therefore, we impose the upper limit of $B=10$.
The hyperparameter tuning for each local model, following the procedure described in Section 3.3, was executed efficiently in parallel using 10 threads. 
Prior to model fitting, all dimensions of the covariates were standardized.

We compare our proposed PPK against several standard baselines in HTE estimation. For scalable GP approximations, we employ Sparse Variational GP (SVGP) \citep{NIPS2005_4491777b,pmlr-v5-titsias09a} and Nearest Neighbor GP (NNGP) \citep{Datta02042016}. 
Furthermore, as representative tree-based methods, we evaluate Bayesian Causal Forest (BCF) \citep{hahn2019bayesianregressiontreemodels}, Generalized Random Forest (GRF) \citep{athey2018generalizedrandomforests}, and Random Forest within the R-learner framework \citep{nie2020quasioracleestimationheterogeneoustreatment}. Due to space limitations, the results for NNGP, GRF, and the R-learner are omitted from the main plots in this section. The comprehensive results, including all comparison methods, are provided in Section S3 of the Supplementary Material. All computations were carried out on a workstation with 12th Gen Intel Core i7-12650H processor (16 logical cores) and 32.0 GB of RAM.

\begin{figure}[htb!]
    \centering
    \includegraphics[width=0.95\linewidth]{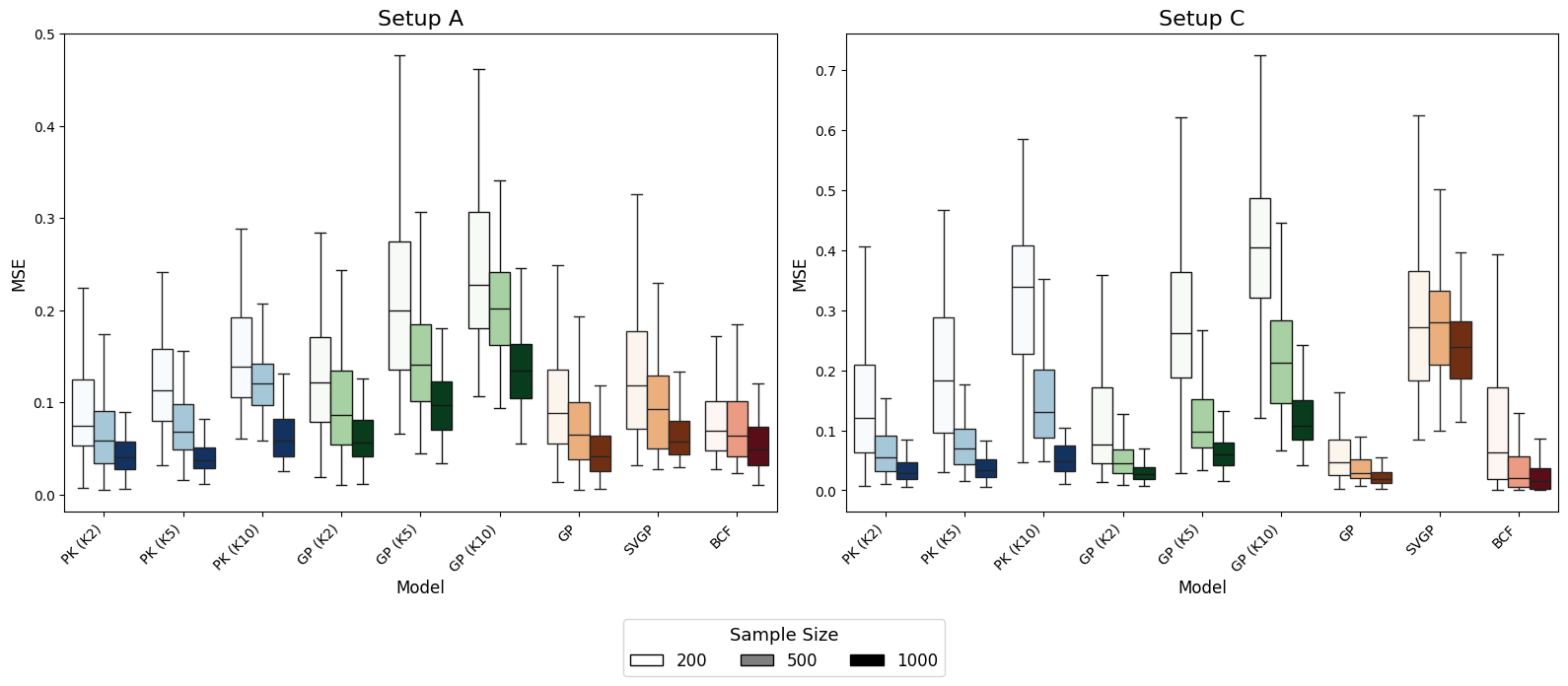}
    \caption{
   Boxplots of mean squared errors (MSEs) for estimating HTE in Setups A and C. 
   Each panel compares PPK, the simple local GP approximation, the baseline global GP, SVGP and BCF across different sample sizes.}
\label{joined_mse_AandC}
\end{figure}

Figure~\ref{joined_mse_AandC} compares the MSEs of the competing methods in Setups A and C.
Note that in almost all cases, PPK outperforms simple local approximation for the same sample size and number of regions.
The only cases in which PPK does not clearly outperform the simple local approximation occur in Setup C with $K=2$.
In $K=2$ case, the number of boundaries between regions is only one, so effect of the \textit{patchwork} scheme is expected to be weaker than in the $K=5$ or $10$ cases. 
Within each divided region, the local model uses fewer observations than the baseline global GP. However, the negative effect of the reduced local sample size appears to be relatively small.
Unlike the baseline global GP, both PPK and the simple local approximation can use region-specific hyperparameters, allowing them to capture locally varying patterns and smoothness.
In contrast, SVGP is designed as a scalable approximation to the global GP model, and in our experiments it did not improve upon the full GP in estimation accuracy. 
These results suggest that the region-specific adaptability of PPK can partly compensate for the loss of information caused by using smaller local samples.

\begin{figure}[htb!]
    \centering
    \includegraphics[width=0.95\linewidth]{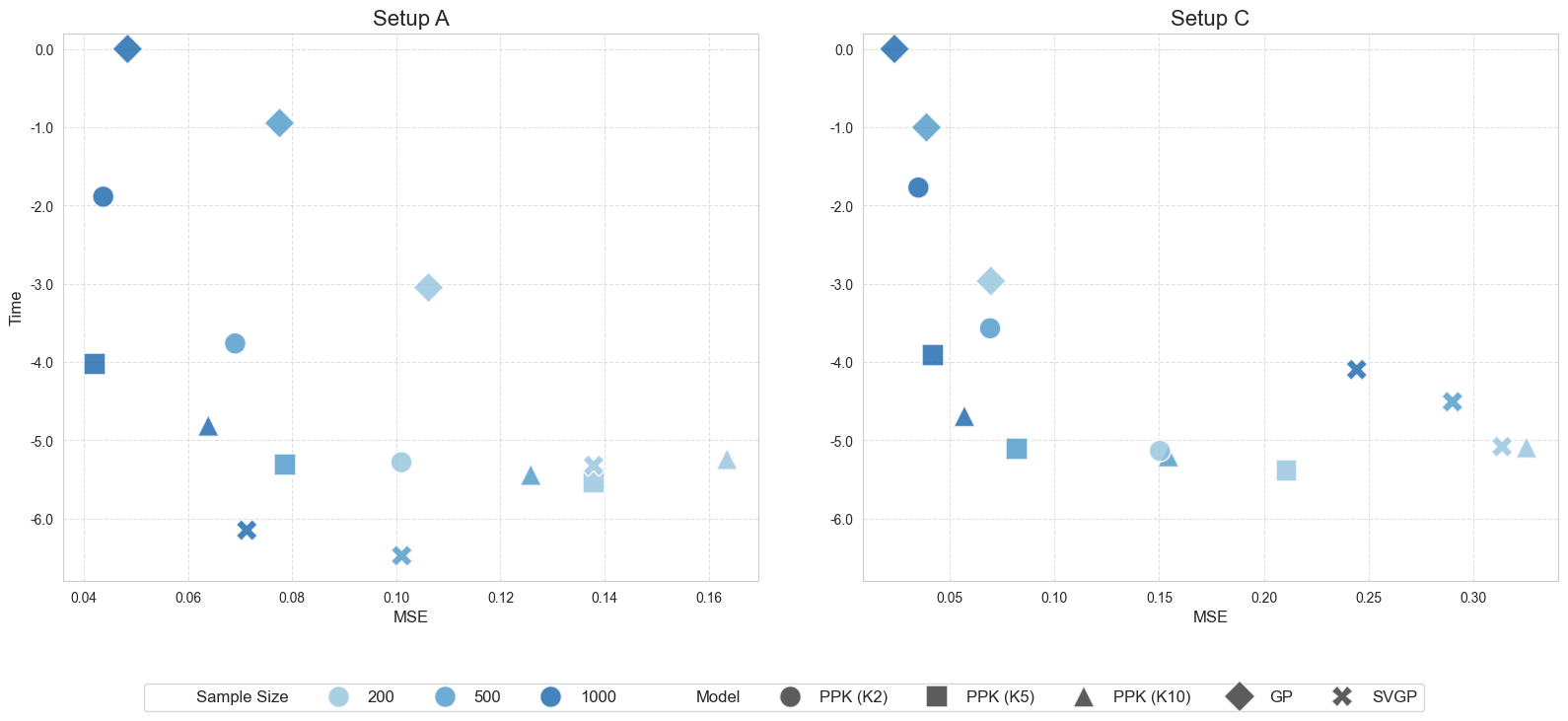}
    \caption{
   Trade-off between log computation time and MSE in Setups A and C. Each point represents one method and sample-size combination.
    }
    \label{joined_time_mse_AandC}
\end{figure}

As shown in Figure~\ref{joined_time_mse_AandC}, the computation time patterns are broadly similar between Setups A and C.
The parallelized hyperparameter tuning contributes the most to the reduction in computation time. For smaller sample sizes, the benefit of increasing the number of partitions is less pronounced due to the overhead associated with parallelization. However, for the larger sample size of $N=1000$ (indicated by the darker plots), the improvement in computation speed achieved by introducing PPK and increasing the number of partitions becomes highly prominent. Furthermore, it is evident that SVGP does not outperform the baseline global GP in estimation accuracy for any sample size considered here.
In contrast, PPK can reduce runtime while maintaining, and in some cases improving, estimation accuracy. 
This appears to be due to the combination of local computation and region-specific hyperparameter tuning.

\begin{figure}[H]
    \centering
    \includegraphics[width=\linewidth]{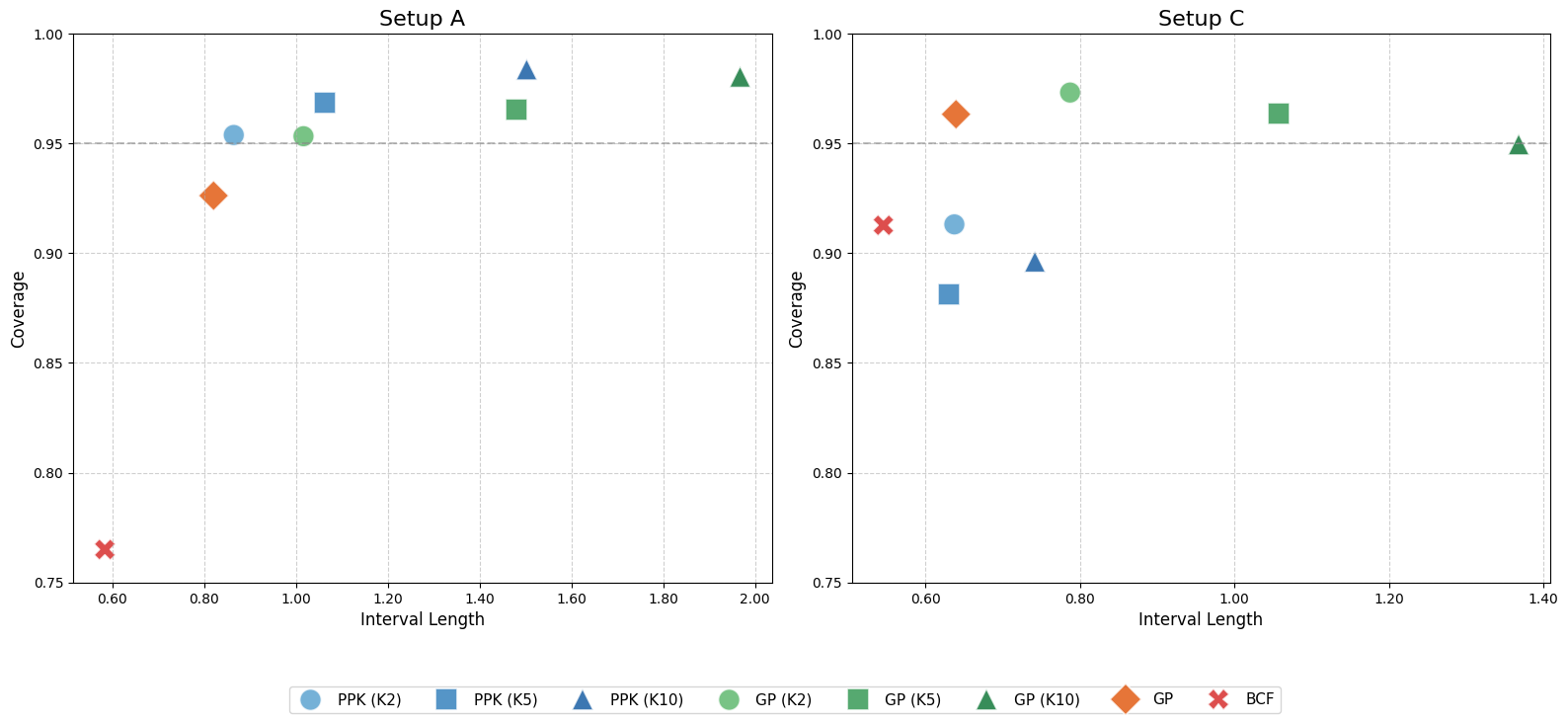}
    \caption{
    Average interval length and empirical coverage probability of 95\% credible intervals in Setups A and C. Each point corresponds to a competing method, and the dashed horizontal line indicates the nominal 95\% coverage level.}
    \label{joined_length_coverage_AandC}
\end{figure}

As discussed in Section 3.5, the proposed method tends to yield narrower credible intervals due to the introduction of pseudo-observations. Indeed, Figure~\ref{joined_length_coverage_AandC} demonstrates that PPK results in shorter interval lengths compared to the simple local approximation, which correspondingly leads to lower empirical coverage rates. 
However, in the present settings, choosing $B$ moderately relative to the regional sample size $N_k$ helps prevent the empirical coverage from becoming substantially lower than the nominal 95\% level.

\begin{figure}[H]
    \centering
    \includegraphics[width=0.95\linewidth]{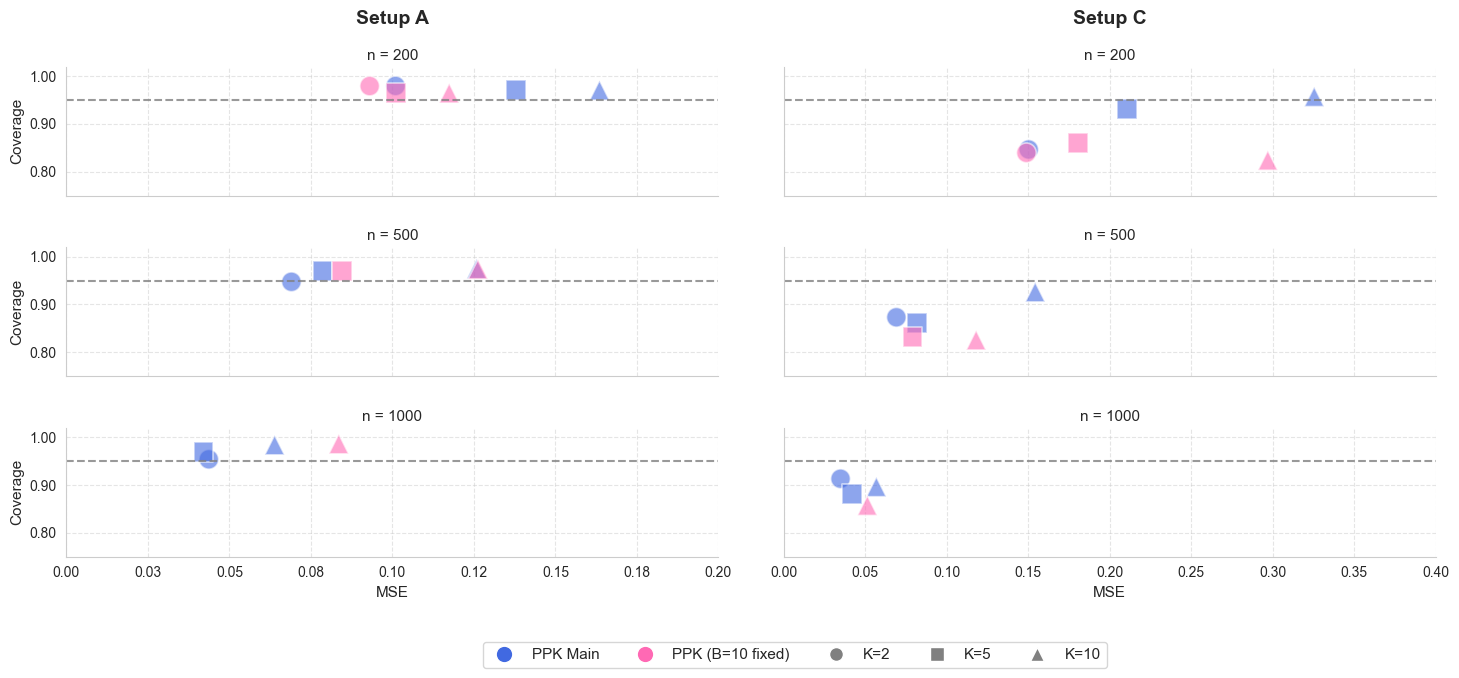}
    \caption{
    Comparison between PPK with the proposed sample-size-adjusted choice of $B$, PPK with a fixed choice $B = 10$ and local GP approximation. The horizontal axis shows
    MSE, and the vertical axis shows empirical coverage probability.
    }
    \label{joined_mse_coverage_with_B10}
\end{figure}

Finally, we compare the PPK model using a fixed number of pseudo-observations ($B=10$) against our proposed PPK model with appropriately adjusted $B$. 
As illustrated in Figure~\ref{joined_mse_coverage_with_B10}, particularly in Setup C, using a large value of $B$, such as $B=10$, can substantially reduce empirical coverage when the regional sample size $N_k$ is small.

\section{Application: The Effect of Smoking on Medical Expenditures}
\subsection{Data}
Here, we consider a real-world analysis that infers the effect of smoking on medical expenditures as an application. This problem has been analyzed in several papers. We use the data from the 1987 National Medical Expenditures Survey (NMES) by \cite{JOHNSON2003135}. The dataset was constructed following the procedure in \cite{Imai01092004}, resulting in a sample size of $N=7752$. The dataset includes the following 10 covariates:

\begin{itemize}
\setlength{\itemsep}{0cm} 
    \item \textbf{age:} age in years at the time of the survey
    \item \textbf{smoke age:} age in years when the individual started smoking
    \item \textbf{gender:} male or female
    \item \textbf{race:} other, black or white
    \item \textbf{marriage status:} married, widowed, divorced, separated, never married
    \item \textbf{education level:} college graduate, some college, high school graduate, other
    \item \textbf{census region:} geographic location, Northeast, Midwest, South, West
    \item \textbf{poverty status:} poor, near poor, low income, middle income, high income
    \item \textbf{seat belt:} does patient regularly use a seat belt when in a car
    \item \textbf{years quit:} how many years since the individual quit smoking.
\end{itemize}
The outcome is the logarithm of annual medical expenditures. To treat the intervention as binary, following \cite{hahn2019bayesianregressiontreemodels},  we assign $T_i=1$ to unit $i$ if their \textit{pack-years} exceed $17$, and $T_i=0$ otherwise. This variable is interpreted as the cumulative amount of smoking; see \cite{Imai01092004} for further details.

In this application, as in the simulation study, we use  PPK, the baseline GP model, a simple local approximation of the baseline and BCF. Due to the larger sample size, we increase the number of regions $K$ to $\{10, 30, 45\}$ and set the number of pseudo-observations at each boundary $B$ to $10$. We performed the tuning method described in Section 3.2.

\subsection{Results}
\begin{figure}[htb!]
    \centering
    \includegraphics[width=0.9\linewidth]{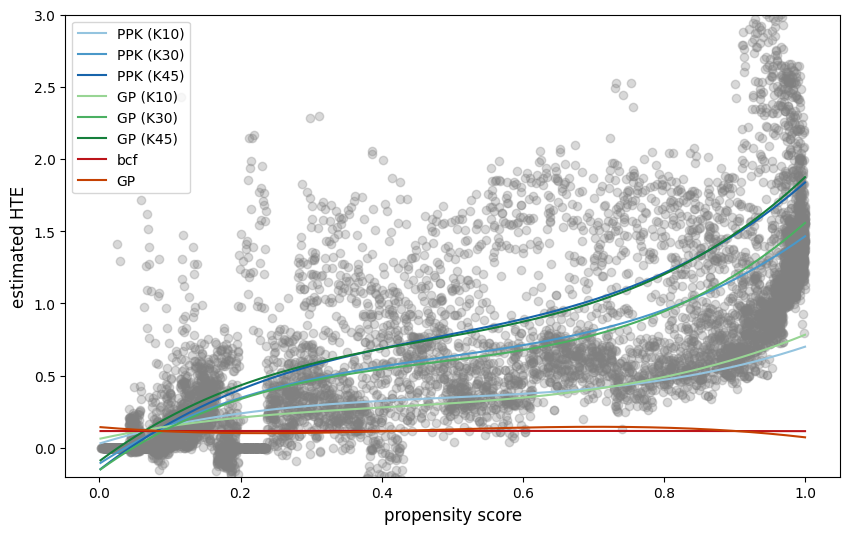}
    \caption{Spline-smoothed trends of estimated HTE as a function of the estimated propensity score for PPK, the simple local GP approximation, the baseline global GP, and BCF. Gray dots indicate individual HTE estimates from PPK with $K=30$.}
    \label{RWD_all_method}
\end{figure}

First, we examine the relationship between the estimated HTE and the propensity scores. 
Figure~ \ref{RWD_all_method} displays the spline-smoothed trends for each method, overlaid with a scatter plot of individual HTE estimates obtained by PPK with $K=30$.
We confirm that the trend of PPK closely matches that of the simple local approximation with the same number of regions. 
In contrast, the spline-smoothed trends for the baseline GP model and BCF are relatively flat over the propensity score range. 
This suggests that, in this application, these global models capture less systematic variation in HTE along the propensity score direction than the partition-based methods.

\begin{figure}[htb!]
    \centering
    \includegraphics[width=1.0\linewidth]{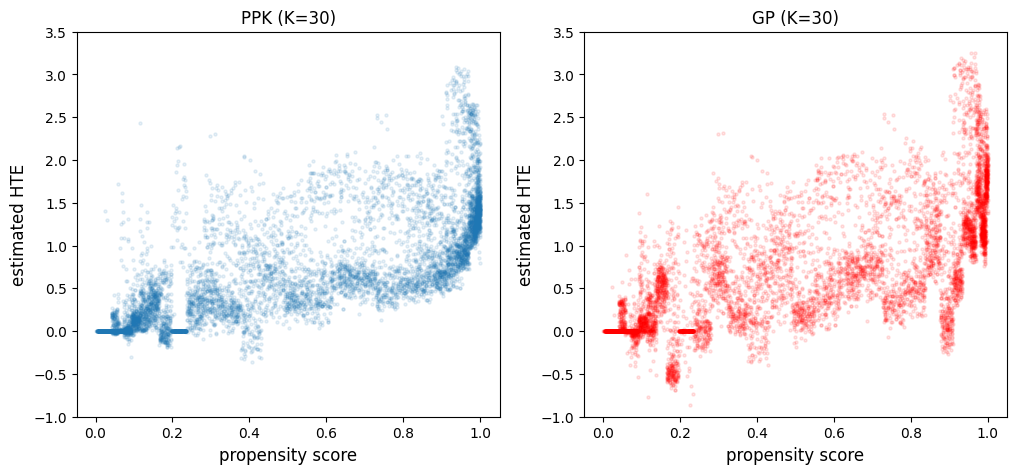}
    \caption{Scatter plots of estimated HTE versus estimated propensity score: PPK with $K=30$ (left) and the simple local GP approximation with $K=30$ (right).
    }
    \label{RWD_K30_parallel}
\end{figure}

\begin{figure}[htb!]
    \centering
    \includegraphics[width=0.5\linewidth]{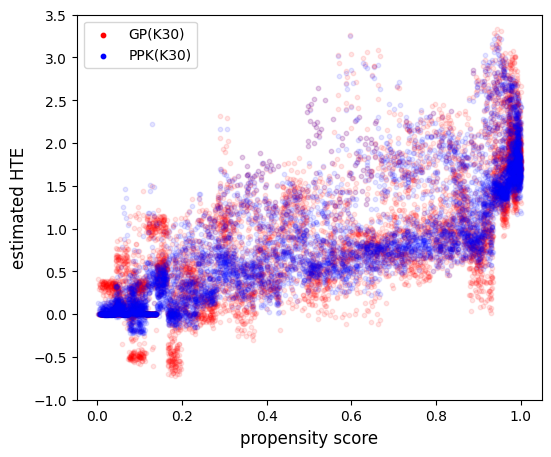}
    \caption{
    Overlaid scatter plots of the estimated HTEs from PPK with $K=30$ and the simple local GP approximation with $K=30$. Blue points indicate PPK estimates, and red points indicate local GP estimates.
    }
    \label{RWD_K30_overlap}
\end{figure}

Second, although the overall spline trends match between PPK and local approximation, there are significant differences at the individual level. Figure \ref{RWD_K30_parallel} shows the scatter plots of the estimates from our proposed method ($K = 30$) and the local approximation ($K = 30$). 
The simple local GP approximation exhibits block-like discontinuities across partition boundaries, whereas PPK connects these blocks more smoothly through the patchwork constraints.
Furthermore, Figure~\ref{RWD_K30_overlap} shows the overlap of these estimates. 
The simple local approximation fits the model using only the limited number of samples within each divided region, making the resulting estimates more unstable within each region.
As seen in Figure~\ref{RWD_K30_overlap}, the estimates from the simple local GP approximation are more dispersed within several propensity score regions, whereas PPK yields smoother transitions across adjacent regions.

Finally, we compare the computation times, as summarized in Table \ref{RWD_time_sec}. 
Due to the data partitioning strategy, all PPK specifications substantially reduce computation time compared with the baseline global GP model.
Note that the experiments were conducted on a CPU with 16 threads, with the maximum number of parallel processes set to 15 for hyperparameter tuning. In the case of $K=10$, all regions can be tuned simultaneously. For $K=45$, the tuning process requires at least three sequential batches (since $45/15 = 3$). However, the reduced sample size within each local region significantly accelerates the convergence of the gradient method during hyperparameter tuning (particularly for $S_\epsilon$). Consequently, PPK with $K=45$ achieves a more than threefold speedup compared to $K=10$, despite the need for sequential batches. 
By contrast, the baseline global GP does not benefit from this partitioning strategy and requires approximately 34 hours in this application.

\begin{table}[htb!]
    \centering
    \caption{
    Computation times in seconds for the medical expenditure application.
    }
    \begin{tabular}{lcccc}
    \hline
     & PPK (K10) & PPK (K30) & PPK (K45) & GP \\ \hline
    Time (sec) & 1740 & 656 & 419 & 125703 \\ \hline
    \end{tabular}
    \label{RWD_time_sec}
\end{table}

\section{Concluding Remarks}
In this paper, we proposed propensity Patchwork Kriging, an efficient HTE estimation method using Gaussian Processes (GP) combined with the Patchwork Kriging technique. Our experiments demonstrated that PPK is computationally more efficient than standard GP approaches, and generally more accurate than simple local approximation methods in the settings considered.
The simulations suggest that partitioning can be particularly effective when the data-generating process exhibits region-specific smoothness or heterogeneity, because PPK allows hyperparameters to be tuned separately within each local region.
A key feature of our approach is the use of estimated propensity scores for data partitioning. 
By adopting this one-dimensional variable, the boundaries between regions become clearer, simplifying the subsequent implementation. Furthermore, PPK can be interpreted as a natural extension of propensity score stratification in causal inference.

Although the propensity score provides a statistically meaningful one-dimensional partitioning variable, continuity along the propensity score is not, in general, a substitute for continuity over the full covariate space. 
If the true HTE varies substantially along covariate directions that are weakly represented by the propensity score, the proposed PPK may not fully capture the relevant structure and may introduce bias. 
In such cases, sensitivity analyses with respect to the number of partitions and boundary pseudo-observations are important diagnostics. 
Extensions that incorporate additional balancing scores or multiple partitioning variables may provide a useful direction for future research.

\section*{Acknowledgement}
We would like to thank two anonymous reviewers for their valuable comments and suggestions, which significantly improved the paper. 
This work is partially supported by JSPS KAKENHI Grant Numbers 24K21420 and 25H00546.

\vspace{1cm}

\bibliographystyle{chicago}
\bibliography{ref}

\newpage
\setcounter{equation}{0}
\setcounter{section}{0}
\setcounter{table}{0}
\setcounter{figure}{0}
\setcounter{page}{1}
\renewcommand{\thesection}{S\arabic{section}}
\renewcommand{\theequation}{S\arabic{equation}}
\renewcommand{\thetable}{S\arabic{table}}
\renewcommand{\thefigure}{S\arabic{figure}}

\vspace{1cm}
\begin{center}
{\LARGE
{\bf Supplementary Material for ``Propensity Patchwork Kriging for Scalable Inference on Heterogeneous Treatment Effects"}
}
\end{center}

\medskip
This Supplementary Material provides details of the derivation of the proposed method, data generation process of simulation study, additional simulation results and the proof of Proposition~1.

\section{Details of Derivation}

\subsection{Prior precision matrix}
In Section~3.4, we provide a detailed derivation of prior precision matrix of $\bm{\Theta}$ in \eqref{prior-precision}. We defined the precision matrix as the inverse of covariance matrix, $\Delta_{\bm{\Theta},\bm{\Theta}} = {\Sigma}^{-1}_{\bm{\Theta},\bm{\Theta}}$, where $\Sigma_{\bm{\Theta},\bm{\Theta}}$ is covariance matrix from  \eqref{prior}. 
To compute the inverse of this block matrix, we use the Schur's complement as follows;
\begin{equation}
\begin{split}
    \label{Schur's complement}
    \begin{bmatrix}
        A & B \\
        C & D
    \end{bmatrix}^{-1} &= 
    \begin{bmatrix}
        M^{-1} & -M^{-1} B D^{-1} \\
        -D^{-1} C M^{-1} & D^{-1}+ D^{-1} C M^{-1}BD^{-1}   \end{bmatrix}\\
    M &= A - BD^{-1} C
\end{split}
\end{equation}
The precision matrix is inverse of $4 \times 4$ block matrix.
\begin{equation*}
\begin{split}
\Delta_{\bm{\Theta}, \bm{\Theta}} &= {\Sigma}^{-1}_{\bm{\Theta}, \bm{\Theta}} 
=  \begin{bmatrix}
        \Sigma_{\bm{\theta}_n, \bm{\theta}_n} & \Sigma_{\bm{\theta}_n, \tilde{\bm{\theta}}_m} & \bm{0} & \Sigma_{\bm{\theta}_n, \bm{\delta}} \\
        \Sigma_{\tilde{\bm{\theta}}_m, \bm{\theta}_n} & \Sigma_{\tilde{\bm{\theta}}_m, \tilde{\bm{\theta}}_m} & \bm{0} & \Sigma_{\tilde{\bm{\theta}}_m, \bm{\delta}} \\
        \bm{0} & \bm{0} & \Sigma_{\bm{f}_n, \bm{f}_n} & \bm{0}\\
        \Sigma_{\bm{\delta}, \bm{\theta}_n} & \Sigma_{\bm{\delta}, \tilde{\bm{\theta}}_m} & \bm{0} & \Sigma_{\bm{\delta}, \bm{\delta}}
    \end{bmatrix}^{-1}
\end{split}
\end{equation*}
To apply the formula \eqref{Schur's complement}, we partition $\Sigma_{\bm{\Theta}, \bm{\Theta}}$ by defining $A$ as the upper-left $3 \times 3$ block, $B$ as the upper-right $3 \times 1$ block, $C$ as the lower-left $1 \times 3$ block, and $D$ as the scalar block $\Sigma_{\bm{\delta}, \bm{\delta}}$.

First, we compute the term $B D^{-1}C$:
\begin{align*}
\begin{split}
    B D^{-1} C &= \begin{bmatrix}
    \Sigma_{\bm{\theta}_n, \bm{\delta}}\\
    \Sigma_{\tilde{\bm{\theta}}_m, \bm{\delta}}\\
    \bm{0}
\end{bmatrix}  \Sigma_{\bm{\delta},\bm{\delta}}^{-1}
\begin{bmatrix}
     \Sigma_{\bm{\delta}, \bm{\theta}_n} & \Sigma_{\bm{\delta}, \tilde{\bm{\theta}}_m} & \bm{0}
\end{bmatrix} \\
 &= \begin{bmatrix}
\Sigma_{\bm{\theta}_n,\bm{\delta}}\Sigma^{-1}_{\bm{\delta},\bm{\delta}}\Sigma_{\bm{\delta}, \bm{\theta}_n} & 
\Sigma_{\bm{\theta}_n,\bm{\delta}}\Sigma^{-1}_{\bm{\delta},\bm{\delta}}\Sigma_{\bm{\delta}, \tilde{\bm{\theta}}_m} & \bm{0} \\
\Sigma_{\tilde{\bm{\theta}}_m, \bm{\delta}}\Sigma^{-1}_{\bm{\delta},\bm{\delta}}\Sigma_{\bm{\delta}, \bm{\theta}_n} & 
\Sigma_{\tilde{\bm{\theta}}_m,\bm{\delta}}\Sigma^{-1}_{\bm{\delta},\bm{\delta}}\Sigma_{\bm{\delta}, \tilde{\bm{\theta}}_m} & \bm{0}\\
\bm{0} & \bm{0} & \bm{0}
 \end{bmatrix}
\end{split}
\end{align*}
Next, we compute $M$ in \eqref{Schur's complement}:
\begin{equation*}
\begin{split}
    M &= A-BD^{-1}C \\ &= 
    \begin{bmatrix}
\Sigma_{\bm{\theta}_n,\bm{\theta}_n} - \Sigma_{\bm{\theta}_n,\bm{\delta}}\Sigma^{-1}_{\bm{\delta},\bm{\delta}}\Sigma_{\bm{\delta}, \bm{\theta}_n}& 
\Sigma_{\bm{\theta}_n,\tilde{\bm{\theta}}_m} - \Sigma_{\bm{\theta}_n,\bm{\delta}}\Sigma^{-1}_{\bm{\delta},\bm{\delta}}\Sigma_{\bm{\delta}, \tilde{\bm{\theta}}_m} & \bm{0} \\
\Sigma_{\tilde{\bm{\theta}}_m, \bm{\theta}_n} - \Sigma_{\tilde{\bm{\theta}}_m, \bm{\delta}}\Sigma^{-1}_{\bm{\delta},\bm{\delta}}\Sigma_{\bm{\delta}, \bm{\theta}_n} & 
\Sigma_{\tilde{\bm{\theta}}_m,\tilde{\bm{\theta}}_m} - \Sigma_{\tilde{\bm{\theta}}_m,\bm{\delta}}\Sigma^{-1}_{\bm{\delta},\bm{\delta}}\Sigma_{\bm{\delta}, \tilde{\bm{\theta}}_m} & \bm{0}\\
\bm{0} & \bm{0} & \Sigma_{f_n, f_n}
 \end{bmatrix}.
\end{split}
\end{equation*}
To compute the inverse of $M$, we use \eqref{Schur's complement} repeatedly. 
Let $A'$ be the upper-left $2 \times 2$ block, $B'$ as $2 \times 1$ ,$C'$ as lower-left $1 \times 2$ block and  $D'$ as $\Sigma_{\bm{f}_n, \bm{f}_n}$. Notice that off-diagonal matrices $B'$ and $C'$ are zero-matrix. Also, let $A''$ through $D''$ denote the blocks of $A'$ 

Because $B'$ and $C'$ are zero-matrices, applying \eqref{Schur's complement} to $M$, $M^{-1}$ is a block diagonal matrix as follows:
\begin{equation*}
\begin{split}
    M^{-1} &= \begin{bmatrix}
        A'^{-1}  & \bm{0} \\
        \bm{0} & D'^{-1}
    \end{bmatrix}
    =\begin{bmatrix}
        M'' & - M''^{-1} B'' C'' & \bm{0} \\
        -D^{-1}C''M''^{-1} & D''^{-1} - D^{-1}C''M''^{-1} B D^{-1} & \bm{0} \\
        \bm{0} & \bm{0} & \Sigma^{-1}_{\bm{f}_n, \bm{f}_n}
    \end{bmatrix}\\
    &\equiv \begin{bmatrix}
        \Delta_{\bm{\theta}_n, \bm{\theta}_n} & \Delta_{\bm{\theta}_n, \tilde{\bm{\theta}}_m} & \bm{0} \\
        \Delta_{\tilde{\bm{\theta}}_m, \bm{\theta}_n} & \Delta_{\tilde{\bm{\theta}}_m, \tilde{\bm{\theta}}_m} & \bm{0} \\
        \bm{0} & \bm{0} & \Delta_{\bm{f}_n, \bm{f}_n}
    \end{bmatrix},
\end{split}
\end{equation*}
where $M'' = A'' - B'' D''^{-1} C''$.

Now we compute the upper-right blocks in the \eqref{Schur's complement} to obtain 
\begin{equation}
\label{MBD}
\begin{split}
    -M^{-1}BD^{-1} 
    &= \begin{bmatrix}
        -(\Delta_{\bm{\theta}_n, \bm{\theta}_n}\Sigma_{\bm{\theta}_n, \bm{\delta}} + \Delta_{\bm{\theta}_n, \tilde{\bm{\theta}}_m}\Sigma_{\tilde{\bm{\theta}}_m, \bm{\delta}})\Sigma^{-1}_{\bm{\delta},\bm{\delta}} \\
        -(\Delta_{\tilde{\bm{\theta}}_m, \bm{\theta}_n}\Sigma_{\bm{\theta}_n, \bm{\delta}} + \Delta_{\tilde{\bm{\theta}}_m, \tilde{\bm{\theta}}_m}\Sigma_{\tilde{\bm{\theta}}_m, \bm{\delta}})\Sigma^{-1}_{\bm{\delta},\bm{\delta}} \\
        \bm{0}
    \end{bmatrix}
    \equiv \begin{bmatrix}
        \Delta_{\bm{\theta}_n, \bm{\delta}} \\ \Delta_{\tilde{\bm{\theta}}_m, \bm{\delta}}\\
        \bm{0}
    \end{bmatrix}
\end{split}
\end{equation}
The lower-left block in \eqref{Schur's complement} is the transpose of \eqref{MBD}, obtained as 
\begin{equation*}
\begin{split}
-&D^{-1}CM^{-1} \\
&= \begin{bmatrix}
        -\Sigma^{-1}_{\bm{\delta},\bm{\delta}}(\Sigma_{\bm{\delta}, \bm{\theta}_n}\Delta_{\bm{\theta}_n, \bm{\theta}_n} + \Sigma_{\bm{\delta}, \tilde{\bm{\theta}}_m}\Delta_{\tilde{\bm{\theta}}_m, \bm{\theta}_n}) &
         -\Sigma^{-1}_{\bm{\delta},\bm{\delta}}(\Sigma_{\bm{\delta}, \bm{\theta}_n}\Delta_{\bm{\theta}_n, \tilde{\bm{\theta}}_m} + \Sigma_{\bm{\delta}, \tilde{\bm{\theta}}_m}\Delta_{\tilde{\bm{\theta}}_m, \tilde{\bm{\theta}}_m}) &
        \bm{0}
    \end{bmatrix} \\
    &\equiv \begin{bmatrix}
        \Delta_{\bm{\delta, \bm{\theta}_n}} & \Delta_{\bm{\delta, \tilde{\bm{\theta}}_m}} & \bm{0}
    \end{bmatrix}
\end{split}
\end{equation*}
The lower-right block is scalar block as follows:
\begin{equation*}
\begin{split}
    D^{-1}+ D^{-1} C M^{-1}BD^{-1} &=
    \Sigma_{\bm{\delta},\bm{\delta}} - (\Delta_{\bm{\delta}, \bm{\theta}_n}\Sigma_{\bm{\theta}_n \bm{\delta}}+ \Delta_{\bm{\delta}, \tilde{\bm{\theta}}_m}\Sigma_{\tilde{\bm{\theta}}_m \bm{\delta}})\Sigma_{\bm{\delta},\bm{\delta}}^{-1}\\
    &\equiv \Delta_{\bm{\delta},\bm{\delta}}
\end{split}
\end{equation*}

\subsection{Joint covariance matrix}
This section details the derivation of $\hat{\Sigma} = \hat{\Delta}^{-1}$, which is the covariance matrix of the joint distribution $p(\bm{\Theta}, \bm{y}_n, \bm{\delta}|\bm{X}_n, \tilde{\bm{X}}_m)$.  We have defined $\hat{\Delta}$ as \eqref{delta-hat}. To compute the inverse matrix, we use Schur's complement \eqref{Schur's complement} repeatedly, as in Section~S1.

From the definition, the precision matrix $\hat{\Delta}$ is a $5 \times 5$ block matrix.
\begin{equation*}
\label{delta-hat_-sumed}
\begin{split}
    \hat{\Delta } = 
    \begin{bmatrix}
        \Delta_{\bm{\theta}_n, \bm{\theta}_n} + s_{\epsilon} \bm{T}_n^2 & \Delta_{\bm{\theta}_n, \tilde{\bm{\theta}}_m} & 
        \bm{T}_n^T s_{\epsilon} & \Delta_{\bm{\theta}_n, \bm{\delta}} & -s_{\epsilon} \bm{T}_n\\
        \Delta_{\tilde{\bm{\theta}}_m, \bm{\theta}_n} & \Delta_{\tilde{\bm{\theta}}_m, \tilde{\bm{\theta}}_m} & \bm{0} &
        \Delta_{\tilde{\bm{\theta}}_m, \bm{\delta}} & \bm{0}\\
         \bm{T}_n^T s_{\epsilon} & \bm{0} &  \Delta_{\bm{f}_n, \bm{f}_n} +  s_{\epsilon}\bm{I}_n  &
         \bm{0} & -s_{\epsilon}\bm{I}_n \\
         \Delta_{\bm{\delta}, \bm{\theta}_n} & \Delta_{\bm{\delta}, \tilde{\bm{\theta}}_m} &\bm{0} & \Delta_{\bm{\delta}, \bm{\delta}}& \bm{0} \\
         -s_{\epsilon}\bm{I}_n \bm{T}_n &  \bm{0} & -s_{\epsilon}\bm{I}_n & \bm{0} & s_{\epsilon} \bm{I}_n
    \end{bmatrix}
\end{split}
\end{equation*}
To apply the formula \eqref{Schur's complement}, we partition the matrix by defining $A$ as the upper-left $4 \times 4$ block, $B$ as the upper-right $4 \times 1$ block, $C$ as the lower-left $1 \times 4$ block, and $D$ as the scalar block $s_{\epsilon} \bm{I}_n$.
Next, we compute $M$ in \eqref{Schur's complement}.
\begin{align*}
    M &= A - B D^{-1}C\\
      &= \begin{bmatrix}
         \Delta_{\bm{\theta}_n, \bm{\theta}_n} & \Delta_{\bm{\theta}_n, \tilde{\bm{\theta}}_m} & \bm{0} & \Delta_{\bm{\theta}_n, \bm{\delta}} \\
         \Delta_{\tilde{\bm{\theta}}_m, \bm{\theta}_n} & \Delta_{\tilde{\bm{\theta}}_m, \tilde{\bm{\theta}}_m} & \bm{0} & \Delta_{\tilde{\bm{\theta}}_m, \bm{\delta}} \\
         \bm{0} & \bm{0} &  \Delta_{\bm{f}_n, \bm{f}_n} & \bm{0} \\
          \Delta_{\bm{\delta}, \bm{\theta}_n} & \Delta_{\bm{\delta}, \tilde{\bm{\theta}}_m} &
          \bm{0} & \Delta_{\bm{\delta},\bm{\delta}}
         \end{bmatrix}
\end{align*}
To compute the inverse matrix of $M$, we use \eqref{Schur's complement} again. By partitioning $M$ and defining $A'$, $B'$, $C'$ and $D'$ similarly. we can compute $M'$ as follows.
\begin{align*}
    M' &=  A' - B' {D'}^{-1}C'\\
       &= \begin{bmatrix}
          \Delta_{\bm{\theta}_n, \bm{\theta}_n} - \Delta_{\bm{\theta}_n, \bm{\delta}} \Delta_{\bm{\delta}, \bm{\delta}}^{-1} \Delta_{\bm{\delta}, \bm{\theta}_n} & 
          \Delta_{\bm{\theta}_n, \tilde{\bm{\theta}}_m} - \Delta_{\bm{\theta}_n, \bm{\delta}} \Delta_{\bm{\delta}, \bm{\delta}}^{-1} \Delta_{\bm{\delta}, \tilde{\bm{\theta}}_m} &
          \bm{0} \\
          \Delta_{\tilde{\bm{\theta}}_m,\bm{\theta}_n} - \Delta_{\tilde{\bm{\theta}}_m, \bm{\delta}} \Delta_{\bm{\delta}, \bm{\delta}}^{-1} \Delta_{\bm{\delta}, \bm{\theta}_n} &
          \Delta_{\tilde{\bm{\theta}}_m, \tilde{\bm{\theta}}_m} - \Delta_{\tilde{\bm{\theta}}_m, \bm{\delta}} \Delta_{\bm{\delta}, \bm{\delta}}^{-1} \Delta_{\bm{\delta}, \tilde{\bm{\theta}}_m} & \bm{0} \\
          \bm{0} & \bm{0} & \Delta_{\bm{f}_n, \bm{f}_n}
          \end{bmatrix}
\end{align*}
We partition $M'$ and define $A''$ , $B''$, $C''$ and $D''$ using the same procedure. We find that the off-diagonal blocks $B''$ and $C''$ are zero-matrices. This simplifies the calculation, as $M'' = A''$ and thus ${M''}^{-1} = {A''}^{-1}$.

Next, we partition  $A''$ (the upper-left block of $2 \times 2$ of $M'$) into its components:
\begin{equation*}
\begin{split}
    A''' &= \Delta_{\bm{\theta}_n, \bm{\theta}_n} - \Delta_{\bm{\theta}_n, \bm{\delta}} \Delta_{\bm{\delta}, \bm{\delta}}^{-1} \Delta_{\bm{\delta}, \bm{\theta}_n}\\
    B''' &= \Delta_{\bm{\theta}_n, \tilde{\bm{\theta}}_m} - \Delta_{\bm{\theta}_n, \bm{\delta}} \Delta_{\bm{\delta}, \bm{\delta}}^{-1} \Delta_{\bm{\delta}, \tilde{\bm{\theta}}_m}\\
    C''' &= \Delta_{\tilde{\bm{\theta}}_m,\bm{\theta}_n} - \Delta_{\tilde{\bm{\theta}}_m, \bm{\delta}} \Delta_{\bm{\delta}, \bm{\delta}}^{-1} \Delta_{\bm{\delta}, \bm{\theta}_n}\\
    D''' &= \Delta_{\tilde{\bm{\theta}}_m, \tilde{\bm{\theta}}_m} - \Delta_{\tilde{\bm{\theta}}_m, \bm{\delta}} \Delta_{\bm{\delta}, \bm{\delta}}^{-1} \Delta_{\bm{\delta}, \tilde{\bm{\theta}}_m}
\end{split}
\end{equation*}
Applying \eqref{Schur's complement}, we can compute ${M''}^{-1}$. In addition, adopting the notation for our target matrix $\hat{\Sigma}$, we define the resulting components as follows:

\begin{equation}
\begin{split}
    \label{A2_M_dd_inv}
    {M''}^{-1} &= \begin{bmatrix}
        {M'''}^{-1} & -{M'''}^{-1}B'''{D'''}^{-1} \\
         -{D'''}^{-1}C'''{M'''}^{-1} & {D'''}^{-1} + {D'''}^{-1}C''' {M'''}^{-1}B'''{D'''}^{-1}
    \end{bmatrix} \\
    &\equiv \begin{bmatrix}
        \hat{\Sigma}_{\bm{\theta}_n, \bm{\theta}_n} & \hat{\Sigma}_{\bm{\theta}_n, \tilde{\bm{\theta}}_m} \\
        \hat{\Sigma}_{\tilde{\bm{\theta}}_m, \bm{\theta}_n} & \hat{\Sigma}_{\tilde{\bm{\theta}}_m, \tilde{\bm{\theta}}_m}
    \end{bmatrix}
\end{split}
\end{equation}
where $M''' = A''' - B''' {D'''}^{-1} C'''$.
Similarly, we can compute ${M'}^{-1}$ and ${M}^{-1}$

\begin{equation*}
\begin{split}
    {M'}^{-1} &= \begin{bmatrix}
        {M''}^{-1} & \bm{0} \\
        \bm{0} & {D''}^{-1}
    \end{bmatrix} 
    = \begin{bmatrix}
        {M''}^{-1} & \bm{0} \\
        \bm{0} & \hat{\Sigma}_{\bm{f}_n, \bm{f}_n}
    \end{bmatrix}
\end{split}
\end{equation*}
where $D'' = \Delta_{\bm{f}_n, \bm{f}_n}$ and we define $\hat{\Sigma}_{\bm{f}_n, \bm{f}_n} \equiv {D''}^{-1}$.

\begin{equation}
\label{A2_MBD_d_inv}
\begin{split}
        -M'^{-1}B'D'^{-1} &= - \begin{bmatrix}
        \hat{\Sigma}_{\bm{\theta}_n, \bm{\theta}_n} & \hat{\Sigma}_{\bm{\theta}_n, \tilde{\bm{\theta}}_m} & \bm{0}\\
        \hat{\Sigma}_{\tilde{\bm{\theta}}_m, \bm{\theta}_n} & \hat{\Sigma}_{\tilde{\bm{\theta}}_m, \tilde{\bm{\theta}}_m} & \bm{0}\\
        \bm{0} &  \bm{0} & \hat{\Sigma}_{\bm{f}_n, \bm{f}_n}
    \end{bmatrix}
        \begin{bmatrix}
            \Delta_{\bm{\theta}_n, \bm{\delta}} \\
            \Delta_{\tilde{\bm{\theta}}_m, \bm{\delta}}\\
            \bm{0}
        \end{bmatrix}
        \Delta^{-1}_{\bm{\delta}, \bm{\delta}} \\
    &= \begin{bmatrix}
        - (\hat{\Sigma}_{\bm{\theta}_n, \bm{\theta}_n}\Delta_{\bm{\theta}_n, \bm{\delta}} + \hat{\Sigma}_{\bm{\theta}_n, \tilde{\bm{\theta}}_m}\Delta_{\tilde{\bm{\theta}}_m, \bm{\delta}})\Delta^{-1}_{\bm{\delta}, \bm{\delta}} \\
        - (\hat{\Sigma}_{\tilde{\bm{\theta}}_m, \bm{\theta}_n}\Delta_{\bm{\theta}_n, \bm{\delta}} + \hat{\Sigma}_{\tilde{\bm{\theta}}_m, \tilde{\bm{\theta}}_m}\Delta_{\tilde{\bm{\theta}}_m, \bm{\delta}})\Delta^{-1}_{\bm{\delta}, \bm{\delta}} \\
        \bm{0}
    \end{bmatrix}
\end{split}
\end{equation}
As before, we define the upper block of \eqref{A2_MBD_d_inv} $\hat{\Sigma}_{\bm{\theta}_n,\bm{\delta}}$ and  the middle block as $\hat{\Sigma}_{\tilde{\bm{\theta}}_n,\bm{\delta}}$. Note that $-D'^{-1}C'M'^{-1}$ is the transpose of \eqref{A2_MBD_d_inv}.

By using result of \eqref{A2_M_dd_inv}, 
\begin{equation*}
\begin{split}
    D'^{-1} + D'^{-1}C'M'^{-1}B'D'^{-1} &= \Delta_{\bm{\delta}, \bm{\delta}}^{-1} - \Delta_{\bm{\delta}, \bm{\delta}}^{-1}(\Delta_{\bm{\delta}, \bm{\theta}_n} \hat{\Sigma}_{\bm{\theta}_n, \bm{\delta}} + \Delta_{\bm{\delta}, \tilde{\bm{\theta}}_m} \hat{\Sigma}_{\tilde{\bm{\theta}}_m, \bm{\delta}}) \Delta_{\bm{\delta}, \bm{\delta}}^{-1}\\
    &\equiv \hat{\Sigma}_{\bm{\delta}, \bm{\delta}}
\end{split}
\end{equation*}
Summarizing these results, we have 
\begin{equation*}
    M^{-1} = \begin{bmatrix}
         \hat{\Sigma}_{\bm{\theta}_n, \bm{\theta}_n} & \hat{\Sigma}_{\bm{\theta}_n, \tilde{\bm{\theta}}_m} & \bm{0} & \hat{\Sigma}_{\bm{\theta}_n, \bm{\delta}} \\
         \hat{\Sigma}_{\tilde{\bm{\theta}}_m, \bm{\theta}_n} & \hat{\Sigma}_{\tilde{\bm{\theta}}_m, \tilde{\bm{\theta}}_m} & \bm{0} & \hat{\Sigma}_{\tilde{\bm{\theta}}_m, \bm{\delta}} \\
         \bm{0} & \bm{0} &  \hat{\Sigma}_{\bm{f}_n, \bm{f}_n} & \bm{0} \\
          \hat{\Sigma}_{\bm{\delta}, \bm{\theta}_n} & \hat{\Sigma}_{\bm{\delta}, \tilde{\bm{\theta}}_m} &
          \bm{0} & \hat{\Sigma}_{\bm{\delta},\bm{\delta}}
         \end{bmatrix}
\end{equation*}
This matrix corresponds to the upper-left $4 \times 4$ block of $\hat{\Sigma}$.
Finally, we compute the remaining blocks of $\hat{\Sigma}$.
\begin{equation}
\label{A2_MBD_inv}
\begin{split}
    -M^{-1}BD^{-1} = \begin{bmatrix}
        \hat{\Sigma}_{\bm{\theta}_n, \bm{\theta}_n}T_n\\
         \hat{\Sigma}_{\tilde{\bm{\theta}}_m, \bm{\theta}_n}T_n\\
          \hat{\Sigma}_{\bm{f}_n, \bm{f}_n}\\
           \hat{\Sigma}_{\bm{\delta}, \bm{\theta}_n}T_n\\
    \end{bmatrix}
    \equiv \begin{bmatrix}
        \hat{\Sigma}_{\bm{\theta}_n, \bm{y}_n}\\
        \hat{\Sigma}_{\tilde{\bm{\theta}}_m, \bm{y}_n}\\
        \hat{\Sigma}_{\bm{f}_n, \bm{y}_n}\\
         \hat{\Sigma}_{\bm{\delta}, \bm{y}_n}
    \end{bmatrix}.
\end{split}
\end{equation}
This matrix corresponds to the upper-right $4 \times 1$ block of $\hat{\Sigma}$. The lower-left block of $\hat{\Sigma}$ is equivalent to the transpose of \eqref{A2_MBD_inv}.
\begin{equation*}
\begin{split}
D^{-1} + D^{-1}CM^{-1}BD^{-1} = s_{\epsilon}^{-1}I_n + T_n\hat{\Sigma}_{\bm{\theta}_n, \bm{y}_n} + \hat{\Sigma}_{\bm{f}_n, \bm{y}_n}
\end{split}
\end{equation*}
This matrix corresponds to the lower-right scalar block of $\hat{\Sigma}$.

\section{Data Generation Process for Experiment}
In this section, we detail data generation process used in Section~4. 
These settings are taken from \citet{nie2020quasioracleestimationheterogeneoustreatment}.
In our simulation study, we set sample size of training data at $N = \{200, 500, 1000 \}$ and test data at $M = 500$.
In all scenarios (Setup A to D), $\bm{X}_i, T_i$ and $Y_i$ are sampled as follows:
\begin{align*}
 Y_i= \theta(\bm{X}_i)T_i + f(\bm{X}_i) + \epsilon_i, \quad 
 T_i|\bm{X}_i \sim {\rm Ber}(e(\bm{X}_i)),
\end{align*}
where $\bm{X}_i \sim p(\bm{x})$ and $\bm{X}_i \in \mathcal{R}^6$.
In \citet{nie2020quasioracleestimationheterogeneoustreatment}, the baseline function $f(\bm{X})$ is formulated as $f(\bm{X}_i) = b(\bm{X}) - 0.5 \ \theta(\bm{X})$.
In each scenario, the formulation of $p(\bm{x}), e(\bm{X}), \theta(\bm{X})$ and $ b(\bm{X})$ are set differently.

\begin{itemize}
\item[-] {\bf (Setup A)} \ \ 
This setting uses smooth $\theta$ and $f$, where both specifications use common variables ($X_1$ and $X_2$), described as 
\begin{align*}
e(\bm{X}) &= \text{trim}_{0.1}\{\sin(\pi X_1 X_2)\}, \quad \theta(\bm{X}) = \frac{1}{2}(X_1 + X_2), \\
b(\bm{X}) &= \sin(\pi X_1 X_2) + 2(X_3 - 0.5)^2 + X_4 + 0.5 X_5,
\end{align*}
where $\text{trim}_{\eta}(x) = \max \{\eta, \min(x, 1 - \eta) \}$.
Here $X_{ik}\sim U(0,1)$, independently for $k=1,\ldots,6$.

\item[-] {\bf (Setup B)} \ \ 
This setting uses constant $e(\bm{X})$, so all units have equivalent probability for assignment.
\begin{align*}
&e(\bm{X}) = 0.5, \ \ \ \ 
\theta(\bm{X})= X_1 + \log(1+e^{X_2}), \\
&b(\bm{X}) = \max\{ X_1 + X_2 , X_3, 0\} + \max\{X_4, X_5\},
\end{align*}
where $\bm{X}\sim \mathcal{N}(\bm{0}, \bm{I})$.

\item[-] {\bf (Setup C)} \ \ 
This setting uses constant function for $\theta(\bm{X})$. In this setting, there is no heterogeneity for treatment effects.
\begin{align*}
&e(\bm{X}) = \frac{1}{1 + e^{X_2 + X_3}}, \ \ \ \ 
\theta(\bm{X}) = 1, \ \ \ \ 
b(\bm{X}) = 2\log(1+e^{X_1+X_2 + X_3}),
\end{align*}
where $\bm{X}\sim \mathcal{N}(\bm{0}, \bm{I})$.

\item[-] {\bf (Setup D)} \ \ 
This setting uses non-differentiable and structurally similar functions for $\theta(\bm{X})$ and $f(\bm{X})$.
\begin{align*}
&e(\bm{X}) = \frac{1}{1 + e^{-X_1  -X_2}}, \ \ \ \ \theta(\bm{X}) = \max\{ X_1 + X_2 + X_3, 0\} - \max\{X_4+ X_5, 0\}, \\
&b(\bm{X}) = \frac{1}{2}(\max\{ X_1 + X_2 + X_3, 0\} - \max\{X_4+ X_5, 0\}),
\end{align*}
where $\bm{X}\sim \mathcal{N}(\bm{0}, \bm{I})$.
\end{itemize}

\section{Additional Experimental Results }

Here we show additional results of the simulation study. 
In Section 4, we investigate Setup A and C. 
In Setup B and D, approximate trends of results are consistent. 

\begin{figure}[H]
    \centering
    \includegraphics[width=1.0\linewidth]{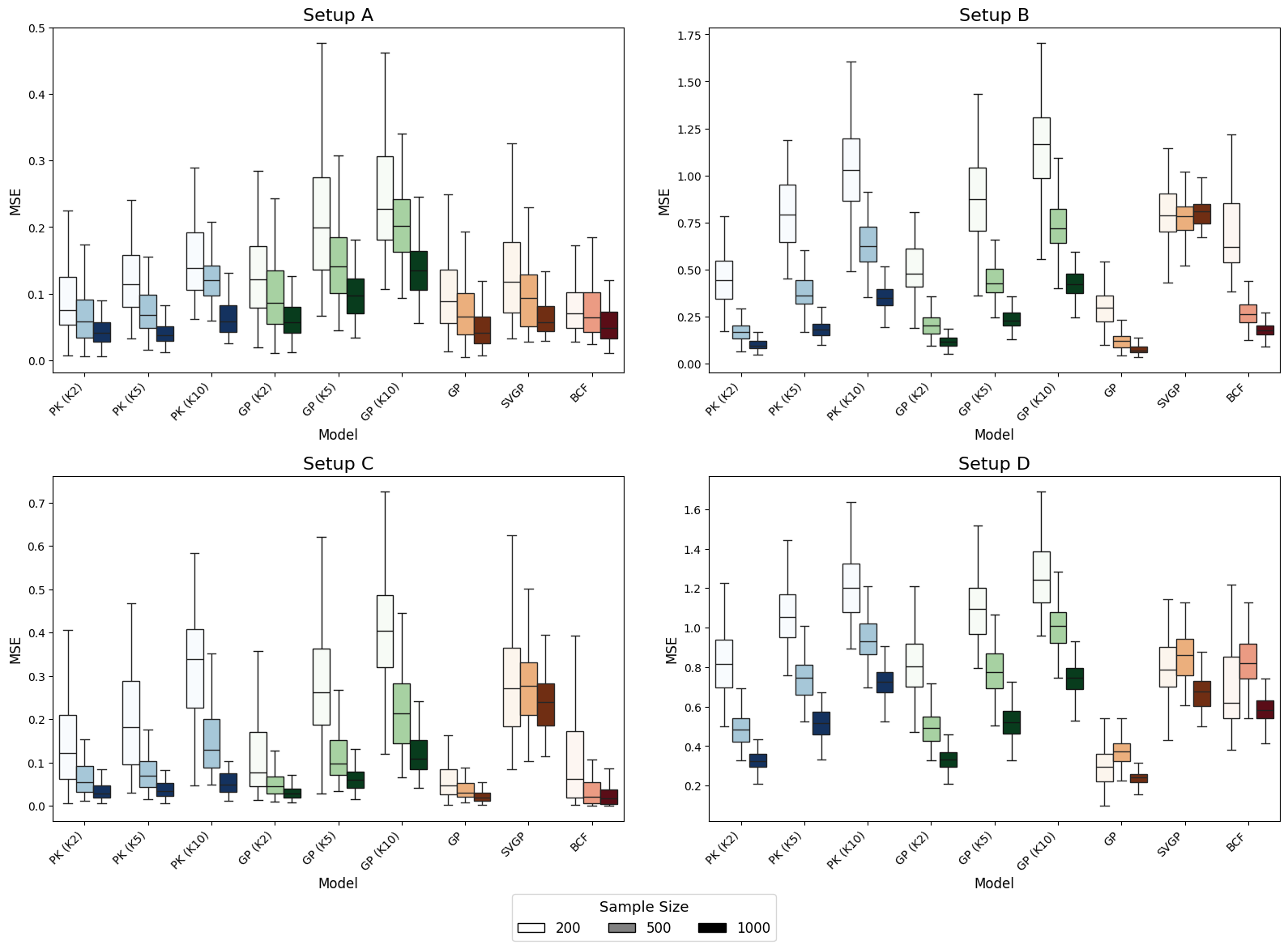}
    \caption{MSEs on all DGP settings}
    \label{all_scenario_mse}
\end{figure}

\begin{figure}[H]
    \centering
    \includegraphics[width=0.9\linewidth]{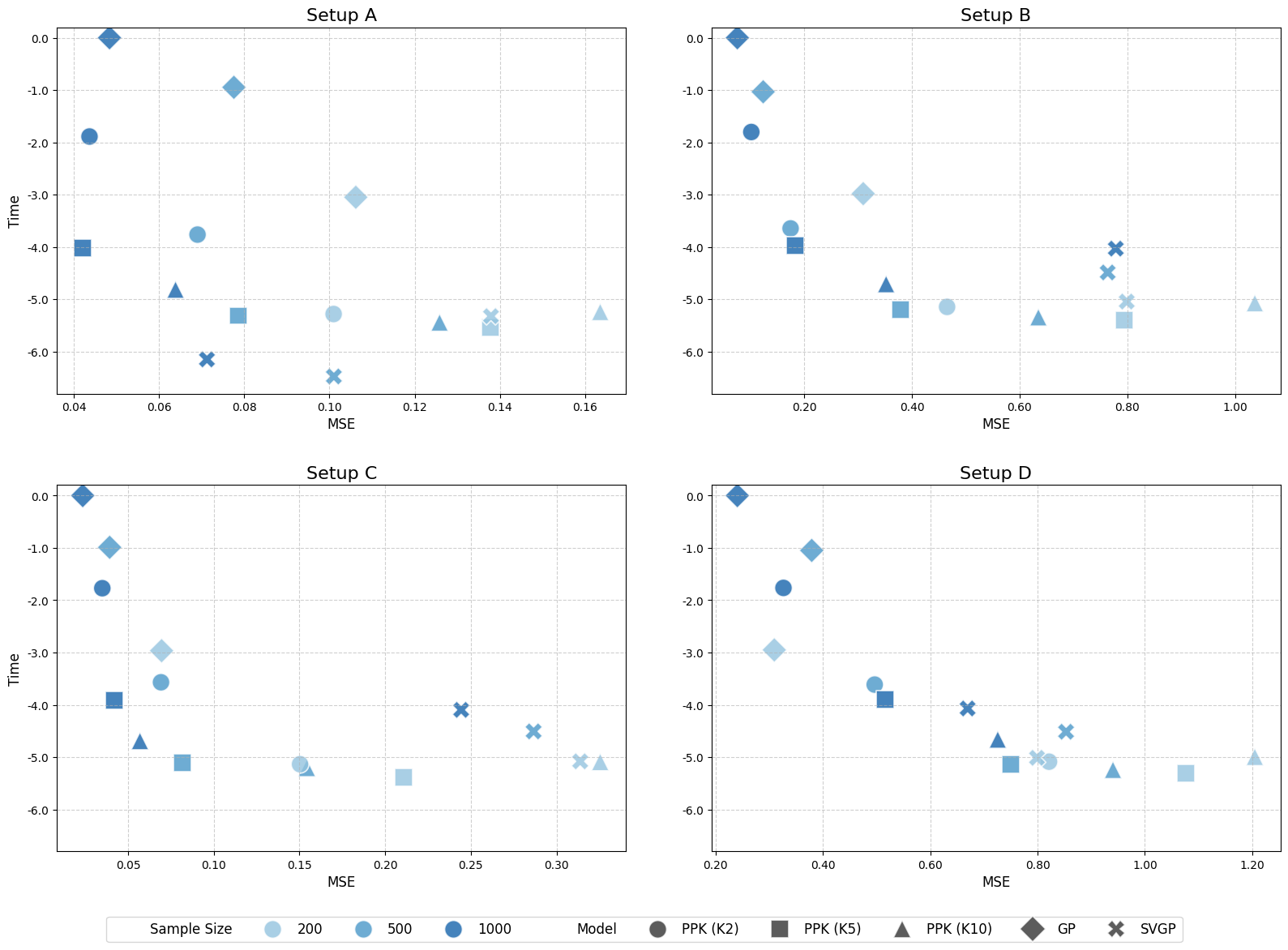}
    \caption{Computation time and MSEs on all DGP settings}
    \label{all_scenario_time_mse}
\end{figure}

\begin{figure}[H]
    \centering
    \includegraphics[width=0.9\linewidth]{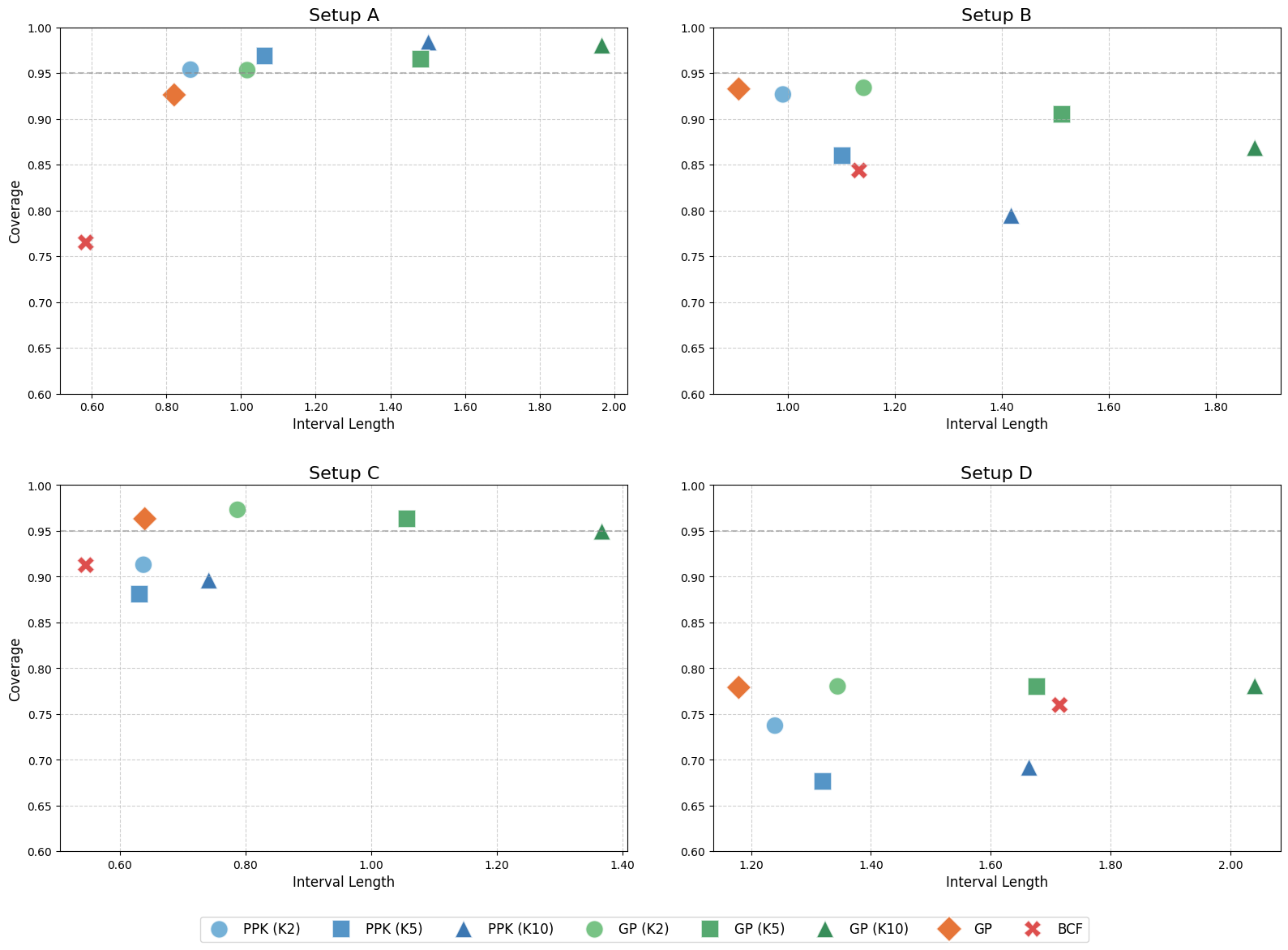}
    \caption{Computation CI length and Coverage on all DGP settings}
    \label{all_scenerio_length_coverage}
\end{figure}

Tables~\ref{metrics_table_A}, \ref{metrics_table_B}, \ref{metrics_table_C} and \ref{metrics_table_D} show the detailed results of the simulation study. 
PPK produces shorter credible interval length (CI length) and smaller empirical coverage rate  (Coverage) than the simple local approximation methods. 
This is likely because the added pseudo-data act as additional constraints in each local model, leading to tighter confidence intervals and, as a result, lower coverage.

\begin{table}[htb!]
\centering
\renewcommand{\arraystretch}{1.0}
\setlength{\tabcolsep}{3pt}
\resizebox{\textwidth}{!}{
\begin{tabular}{l|cccc|cccc|cccc}
\hline
& \multicolumn{4}{c|}{$N = 200$}
& \multicolumn{4}{c|}{$N = 500$}  
& \multicolumn{4}{c}{$N = 1000$} \\ 

\cline{2-13}
         & MSE & CI length & Coverage & Bias & MSE & CI length & Coverage & Bias & MSE & CI length & Coverage & Bias \\
\hline
PPK (K2)  & 0.10 & 1.59 & 0.98 & -0.13 & 0.07 & 1.12 & 0.95 & -0.14 & 0.04 & 0.87 & 0.95 & -0.11 \\
PPK (K5)  & 0.14 & 2.24 & 0.97 & -0.06 & 0.08 & 1.53 & 0.97 & -0.08 & 0.04 & 1.06 & 0.97 & -0.08 \\
PPK (K10) & 0.16 & 2.57 & 0.97 & 0.00 & 0.13 & 2.14 & 0.98 & -0.04 & 0.06 & 1.50 & 0.98 & -0.04 \\
\hline
GP (K2)   & 0.14 & 1.81 & 0.98 & -0.13 & 0.10 & 1.31 & 0.95 & -0.15 & 0.06 & 1.02 & 0.95 & -0.12 \\
GP (K5)   & 0.23 & 2.51 & 0.96 & -0.06 & 0.15 & 1.89 & 0.96 & -0.10 & 0.10 & 1.48 & 0.97 & -0.09 \\
GP (K10)  & 0.25 & 2.90 & 0.97 & -0.01 & 0.21 & 2.40 & 0.96 & -0.04 & 0.14 & 1.97 & 0.98 & -0.05 \\
\hline
GP        & 0.11 & 1.41 & 0.97 & -0.17 & 0.08 & 1.02 & 0.92 & -0.16 & 0.05 & 0.82 & 0.93 & -0.12 \\
NNGP      & 0.14 & 1.99 & 1.00 & -0.25 & 0.10 & 1.75 & 1.00 & -0.23 & 0.07 & 1.55 & 1.00 & -0.20 \\
SVGP      & 0.14 & 2.85 & 1.00 & -0.22 & 0.10 & 2.68 & 1.00 & -0.20 & 0.07 & 2.67 & 1.00 & -0.16 \\
\hline
BCF       & 0.09 & 0.79 & 0.80 & -0.14 & 0.08 & 0.63 & 0.71 & -0.18 & 0.06 & 0.58 & 0.77 & -0.15 \\
GRF       & 0.09 & 1.03 & 1.00 & -0.02 & 0.07 & 0.87 & 1.00 & -0.02 & 0.05 & 0.83 & 1.00 & 0.01 \\
R-learner & 0.35 & 1.58 & 0.96 & 0.27 & 0.27 & 1.42 & 0.97 & 0.28 & 0.21 & 1.33 & 0.99 & 0.21 \\
\hline
\end{tabular}
}
\caption{Performance metrics for Setup A}
\label{metrics_table_A}
\end{table}

\begin{table}[htb!]
\centering
\renewcommand{\arraystretch}{1.0}
\setlength{\tabcolsep}{3pt}
\resizebox{\textwidth}{!}{
\begin{tabular}{l|cccc|cccc|cccc}
\hline
& \multicolumn{4}{c|}{$N = 200$}
& \multicolumn{4}{c|}{$N = 500$}  
& \multicolumn{4}{c}{$N = 1000$} \\ 

\cline{2-13}
         & MSE & CI length & Coverage & Bias & MSE & CI length & Coverage & Bias & MSE & CI length & Coverage & Bias \\
\hline
PPK (K2)  & 0.47 & 1.81 & 0.83 & 0.03 & 0.18 & 1.26 & 0.91 & 0.03 & 0.10 & 0.99 & 0.93 & 0.00 \\
PPK (K5)  & 0.79 & 2.20 & 0.78 & 0.08 & 0.38 & 1.54 & 0.81 & 0.05 & 0.18 & 1.10 & 0.86 & 0.01 \\
PPK (K10) & 1.04 & 2.51 & 0.77 & 0.17 & 0.64 & 2.01 & 0.80 & 0.09 & 0.35 & 1.42 & 0.80 & 0.04 \\
\hline
GP (K2)   & 0.52 & 2.00 & 0.84 & 0.02 & 0.21 & 1.46 & 0.92 & 0.01 & 0.12 & 1.14 & 0.93 & -0.01 \\
GP (K5)   & 0.87 & 2.46 & 0.82 & 0.10 & 0.44 & 1.90 & 0.86 & 0.04 & 0.24 & 1.51 & 0.91 & 0.00 \\
GP (K10)  & 1.14 & 2.84 & 0.81 & 0.17 & 0.72 & 2.30 & 0.83 & 0.10 & 0.42 & 1.87 & 0.87 & 0.03 \\
\hline
GP        & 0.31 & 1.66 & 0.89 & -0.01 & 0.12 & 1.18 & 0.94 & -0.01 & 0.08 & 0.91 & 0.93 & -0.01 \\
NNGP      & 0.37 & 2.60 & 0.97 & -0.01 & 0.17 & 2.27 & 0.99 & -0.05 & 0.12 & 2.07 & 1.00 & -0.07 \\
SVGP      & 0.80 & 3.30 & 0.94 & 0.09 & 0.76 & 3.32 & 0.95 & 0.03 & 0.78 & 3.30 & 0.94 & -0.06 \\
\hline
BCF       & 0.69 & 1.56 & 0.69 & 0.00 & 0.27 & 1.40 & 0.84 & 0.01 & 0.18 & 1.13 & 0.84 & 0.00 \\
GRF       & 0.57 & 1.23 & 0.86 & -0.01 & 0.28 & 1.07 & 0.94 & 0.02 & 0.17 & 0.95 & 0.97 & 0.01 \\
R-learner & 0.77 & 1.95 & 0.89 & 0.17 & 0.54 & 1.74 & 0.93 & 0.10 & 0.44 & 1.61 & 0.94 & 0.19 \\
\hline
\end{tabular}
}
\caption{Performance metrics for Setup B}
\label{metrics_table_B}
\end{table}

\begin{table}[htb!]
\centering
\renewcommand{\arraystretch}{1.0}
\setlength{\tabcolsep}{3pt}
\resizebox{\textwidth}{!}{
\begin{tabular}{l|cccc|cccc|cccc}
\hline
& \multicolumn{4}{c|}{$N = 200$}
& \multicolumn{4}{c|}{$N = 500$}  
& \multicolumn{4}{c}{$N = 1000$} \\ 

\cline{2-13}
         & MSE & CI length & Coverage & Bias & MSE & CI length & Coverage & Bias & MSE & CI length & Coverage & Bias \\
\hline
PPK (K2)  & 0.15 & 1.15 & 0.85 & 0.26 & 0.07 & 0.81 & 0.87 & 0.16 & 0.04 & 0.64 & 0.91 & 0.10 \\
PPK (K5)  & 0.21 & 1.58 & 0.93 & 0.22 & 0.08 & 0.85 & 0.86 & 0.12 & 0.04 & 0.63 & 0.88 & 0.10 \\
PPK (K10) & 0.33 & 2.06 & 0.96 & 0.31 & 0.15 & 1.33 & 0.93 & 0.15 & 0.06 & 0.74 & 0.90 & 0.08 \\
\hline
GP (K2)   & 0.12 & 1.44 & 0.96 & 0.14 & 0.06 & 1.01 & 0.98 & 0.05 & 0.03 & 0.79 & 0.97 & 0.04 \\
GP (K5)   & 0.30 & 2.06 & 0.96 & 0.23 & 0.12 & 1.38 & 0.96 & 0.10 & 0.07 & 1.06 & 0.96 & 0.06 \\
GP (K10)  & 0.41 & 2.56 & 0.97 & 0.31 & 0.22 & 1.82 & 0.96 & 0.15 & 0.12 & 1.37 & 0.95 & 0.09 \\
\hline
GP        & 0.07 & 1.15 & 0.97 & 0.08 & 0.04 & 0.83 & 0.97 & 0.02 & 0.02 & 0.64 & 0.96 & 0.03 \\
NNGP      & 0.11 & 2.15 & 1.00 & 0.05 & 0.08 & 1.86 & 1.00 & 0.02 & 0.06 & 1.64 & 1.00 & 0.02 \\
SVGP      & 0.31 & 3.04 & 1.00 & -0.31 & 0.29 & 2.97 & 1.00 & -0.31 & 0.24 & 2.96 & 1.00 & -0.21 \\
\hline
BCF       & 0.12 & 1.04 & 0.83 & 0.25 & 0.04 & 0.72 & 0.92 & 0.07 & 0.03 & 0.55 & 0.91 & 0.06 \\
GRF       & 0.12 & 1.10 & 1.00 & 0.12 & 0.08 & 0.93 & 1.00 & 0.08 & 0.06 & 0.86 & 1.00 & 0.07 \\
R-learner & 0.72 & 2.01 & 0.90 & 0.43 & 0.50 & 1.71 & 0.94 & 0.39 & 0.45 & 1.64 & 0.93 & 0.29 \\
\hline
\end{tabular}
}
\caption{Performance metrics for Setup C}
\label{metrics_table_C}
\end{table}

\begin{table}[htb!]
\centering
\renewcommand{\arraystretch}{1.0}
\setlength{\tabcolsep}{3pt}
\resizebox{\textwidth}{!}{
\begin{tabular}{l|cccc|cccc|cccc}
\hline
& \multicolumn{4}{c|}{$N = 200$}
& \multicolumn{4}{c|}{$N = 500$}  
& \multicolumn{4}{c}{$N = 1000$} \\ 

\cline{2-13}
         & MSE & CI length & Coverage & Bias & MSE & CI length & Coverage & Bias & MSE & CI length & Coverage & Bias \\
\hline
PPK (K2)  & 0.82 & 1.98 & 0.75 & 0.00 & 0.50 & 1.51 & 0.75 & 0.00 & 0.33 & 1.24 & 0.74 & 0.02 \\
PPK (K5)  & 1.08 & 2.27 & 0.72 & -0.04 & 0.75 & 1.75 & 0.71 & -0.03 & 0.52 & 1.32 & 0.68 & 0.00 \\
PPK (K10) & 1.21 & 2.59 & 0.75 & -0.06 & 0.94 & 2.13 & 0.73 & -0.07 & 0.73 & 1.66 & 0.69 & -0.02 \\
\hline
GP (K2)   & 0.82 & 2.11 & 0.78 & 0.01 & 0.50 & 1.65 & 0.79 & 0.01 & 0.33 & 1.35 & 0.78 & 0.02 \\
GP (K5)   & 1.11 & 2.54 & 0.77 & -0.06 & 0.78 & 2.06 & 0.77 & -0.06 & 0.53 & 1.68 & 0.78 & -0.02 \\
GP (K10)  & 1.26 & 2.91 & 0.79 & -0.08 & 1.02 & 2.41 & 0.77 & -0.09 & 0.74 & 2.04 & 0.78 & -0.04 \\
\hline
GP        & 0.31 & 1.66 & 0.89 & -0.01 & 0.38 & 1.45 & 0.79 & 0.05 & 0.24 & 1.18 & 0.78 & 0.04 \\
NNGP      & 0.37 & 2.60 & 0.97 & -0.01 & 0.51 & 2.31 & 0.92 & 0.07 & 0.39 & 2.12 & 0.93 & 0.05 \\
SVGP      & 0.80 & 3.30 & 0.94 & 0.09 & 0.85 & 3.16 & 0.93 & 0.23 & 0.67 & 3.11 & 0.95 & 0.23 \\
\hline
BCF       & 0.69 & 1.56 & 0.69 & 0.00 & 0.84 & 1.84 & 0.72 & 0.02 & 0.59 & 1.72 & 0.76 & 0.04 \\
GRF       & 1.25 & 1.23 & 0.73 & 0.09 & 0.96 & 1.14 & 0.77 & 0.09 & 0.78 & 1.06 & 0.80 & 0.09 \\
R-learner & 1.57 & 1.87 & 0.76 & 0.45 & 1.53 & 1.64 & 0.74 & 0.48 & 1.42 & 1.54 & 0.75 & 0.50 \\
\hline
\end{tabular}
}
\caption{Performance metrics for Setup D}
\label{metrics_table_D}
\end{table}

\section{Proof of Proposition~1}

We provide regularity conditions and a proof sketch for Proposition 1 in the main text.
The argument is based on the posterior consistency result for the GP-based partially linear model established by \cite{horii2023uncertaintyquantificationheterogeneoustreatment}.
Let $\Pi_{\mathrm{LA},N}(\cdot\mid \boldsymbol{y}_N)$ denote the posterior distribution obtained from the local GP partially linear models without imposing the patchwork constraints, and let $\Pi_{\mathrm{PPK},N}(\cdot\mid \boldsymbol{y}_N,\boldsymbol{\delta}=\boldsymbol{0})$ denote the PPK posterior.
We first note that the PPK posterior can be written as a reweighted version of the
local posterior. 
Indeed, the joint density of the observed outcomes and the boundary constraints can be
decomposed as
$$
p(\boldsymbol{y}_N,\boldsymbol{\delta}=\boldsymbol{0}\mid \tau)
=
p(\boldsymbol{\delta}=\boldsymbol{0}\mid \boldsymbol{y}_N,\tau)
p(\boldsymbol{y}_N\mid \tau).
$$
Thus, the PPK posterior satisfies
\begin{equation}\label{eq:PPK-pos}
\Pi_{\mathrm{PPK},N}(d\tau\mid \boldsymbol{y}_N,\boldsymbol{\delta}=\boldsymbol{0})
=\frac{q_N(\tau)\Pi_{\mathrm{LA},N}(d\tau\mid \boldsymbol{y}_N)}{\int q_N(\tau)\Pi_{\mathrm{LA},N}(d\tau\mid \boldsymbol{y}_N)},
\end{equation}
where $q_N(\tau)=p(\boldsymbol{\delta}=\boldsymbol{0}\mid \boldsymbol{y}_N,\tau)$ is the Gaussian boundary conditioning factor induced by the patchwork constraints.

\begin{enumerate}
\item[(A1)]
The covariate space $\mathcal{X}$ is compact, and the true outcome regression functions admit the partially linear representation
$Y=f_0(X)+T\tau_0(X)+\varepsilon$, where $f_0$ and $\tau_0$ are globally defined functions on $\mathcal{X}$ satisfying the regularity conditions required for posterior consistency of the GP partially linear model of \citet{horii2023uncertaintyquantificationheterogeneoustreatment}. 
These include smoothness and support conditions for $\tau_0$ and $f_0$

\item[(A2)] 
The covariate space is partitioned into a fixed number of regions $\Omega_1,\ldots,\Omega_K$ according to the propensity score, where $K$ does not depend on $N$. 
Each region has positive probability under $P_X$, so that $N_k\to\infty$ for all $k=1,\ldots,K$ as $N\to\infty$, where $N_k$ is the number of observations in $\Omega_k$.

\item[(A3)]
The number of pseudo-observations per boundary $B$ is fixed, and the covariance matrices associated with the patchwork constraints are uniformly non-degenerate. 
Moreover, the Gaussian boundary conditioning factor $q_N(\tau)=p(\boldsymbol\delta=\boldsymbol 0\mid \boldsymbol y_N,\tau)$
is bounded in the sense that there exist constants $0<c<C<\infty$ and a neighborhood $\mathcal{U}_N$ of $\tau_0$ such that, with $P_0$-probability tending to one, $q_N(\tau)\le Cq_N(\tau_0)$ for all $\tau$, and $q_N(\tau)\ge cq_N(\tau_0)$ for all $\tau\in\mathcal{U}_N$.
\end{enumerate}

For each region $\Omega_k$, define the local restrictions of the global functions by $f_{0k}(x)=f_0(x)$ and $\tau_{0k}(x)=\tau_0(x)$ for $x\in\Omega_k$.
These are not separate region-specific target functions, but the restrictions of the common global functions $f_0$ and $\tau_0$ to the $k$th region.
Under (A1), the local functions $f_{0k}$ and $\tau_{0k}$ inherit the same regularity from the global functions $f_0$ and $\tau_0$. 
Therefore, the posterior consistency result for the GP partially linear model applies within each fixed region $\Omega_k$, since $N_k\to\infty$ by (A2).
Condition (A3) is a regularity condition to ensure that this additional finite-dimensional factor does not dominate the likelihood contribution from the observed data.
Since both $K$ and $B$ are fixed, the number of boundary constraints is $B(K-1)=O(1)$.

We now prove Proposition~\ref{prp:ppk_consistency}.
Let $\Pi_{{\rm LA},N}(\cdot| \boldsymbol{y}_N)$ denote the product of the local GP posteriors over the $K$ regions.
By (A1), the local GP posterior in each region is consistent for $\tau_{0k}$.
Since $K$ is fixed and $N_k\to\infty$ for all $k$ by (A2), the product local posterior is consistent for $\tau_0$ in $L_2(P_X)$.
That is, for every $\varepsilon>0$, $\Pi_{{\rm LA},N}\{\tau:\|\tau-\tau_0\|_{L_2(P_X)}>\varepsilon\mid \boldsymbol{y}_N\}\to 0$ in $P_0$-probability.

It remains to show that conditioning on the patchwork constraints does not change this
limiting concentration behavior.
Using the reweighting representation (\ref{eq:PPK-pos}), we have
$$
\Pi_{\mathrm{PPK},N}(A_\varepsilon\mid \boldsymbol{y}_N,\boldsymbol{\delta}=\boldsymbol{0})
=
\frac{
\int_{A_\varepsilon} q_N(\tau)\Pi_{\mathrm{LA},N}(d\tau\mid \boldsymbol{y}_N)
}{
\int q_N(\tau)\Pi_{\mathrm{LA},N}(d\tau\mid \boldsymbol{y}_N)
},
$$
where $A_\varepsilon=\{\tau:\|\tau-\tau_0\|_{L_2(P_X)}>\varepsilon\}$.
Let $\mathcal{U}_N$ be the consistency neighborhood of $\tau_0$ in (A3), so that
$\Pi_{\mathrm{LA},N}(\mathcal{U}_N\mid \boldsymbol{y}_N)\to 1$ in $P_0$-probability.
By (A3), the numerator is bounded above by
$$
\int_{A_\varepsilon} q_N(\tau)\Pi_{\mathrm{LA},N}(d\tau\mid \boldsymbol{y}_N)
\le
Cq_N(\tau_0)\Pi_{\mathrm{LA},N}(A_\varepsilon\mid \boldsymbol{y}_N).
$$
Similarly, the denominator is bounded below by
$$
\int q_N(\tau)\Pi_{\mathrm{LA},N}(d\tau\mid \boldsymbol{y}_N)
\ge
\int_{\mathcal{U}_N} q_N(\tau)\Pi_{\mathrm{LA},N}(d\tau\mid \boldsymbol{y}_N)
\ge
cq_N(\tau_0)\Pi_{\mathrm{LA},N}(\mathcal{U}_N\mid \boldsymbol{y}_N).
$$
Therefore,
$$
\Pi_{\mathrm{PPK},N}(A_\varepsilon
\mid \boldsymbol{y}_N,\boldsymbol{\delta}=\boldsymbol{0})
\le
\frac{
C\Pi_{\mathrm{LA},N}(A_\varepsilon\mid \boldsymbol{y}_N)
}{
c\Pi_{\mathrm{LA},N}(\mathcal{U}_N\mid \boldsymbol{y}_N)
}.
$$
Since $\Pi_{\mathrm{LA},N}(\mathcal{U}_N\mid \boldsymbol{y}_N)\to 1$ and
$\Pi_{\mathrm{LA},N}(A_\varepsilon\mid \boldsymbol{y}_N)\to 0$ in $P_0$-probability, we obtain
$$
\Pi_{\mathrm{PPK},N}(A_\varepsilon
\mid \boldsymbol{y}_N,\boldsymbol{\delta}=\boldsymbol{0})\to 0
$$
in $P_0$-probability.
This completes the proof of Proposition~\ref{prp:ppk_consistency}.

\end{document}